\documentclass[rmp,10pt,floatfix,superscriptaddress,nofootinbib,onecolumn]{revtex4}
\usepackage{amsmath,amsfonts,amssymb,amsthm,graphicx}

\usepackage{bm}
\usepackage{xy}
    \xyoption{matrix}
    \xyoption{arrow}

\usepackage{mathrsfs}

\def\f#1#2{\frac{#1}{#2}}
\def\te#1{\text{#1}}
\def\tilde#1{\widetilde{#1}}

\newcommand{\norm}[1]{\left\| #1 \right\|}
\newcommand{\nrm}[1]{\| #1 \|}

\def\ket#1{\left| #1 \right\rangle}
\def\bra#1{\left\langle #1 \right|}

\newcommand{\lrp}[1]{\left( #1 \right)}
\newcommand{\lrpBig}[1]{\Big( #1 \Big)}
\newcommand{\lrb}[1]{\left[ #1 \right]}
\newcommand{\lrpbig}[1]{\big( #1 \big)}

\newcommand{\threemat}[1]{ \begin{pmatrix} #1 \end{pmatrix} }

\def\({\left(}
\def\){\right)}
\def\<{\left\langle}
\def\>{\right\rangle}

\newcommand{\Tr}{\operatorname{Tr}}

\def\R{\mathbb{R}}

\def\C{\mathbb{C}}
\def\bH{\mathbb{H}}
\def\bO{\mathbb{O}}

\def\eps{\epsilon}

\def\la{\lambda}
\def\La{\Lambda}
\def\de{\delta}
\def\De{\Delta}

\def\m{\mu}

\def\al{\alpha}
\def\om{\omega}

\def\brho{{\bm \rho}}

\def\g{\mathfrak{g}}
\def\fN{\mathfrak{n}}
\def\frakm{\mathfrak{m}}

\def\cV{\mathcal{V}}

\def\cH{\mathcal{H}}

\def\cD{\mathcal{D}}

\def\cA{\mathcal{A}}

\def\cU{\mathcal{U}}

\def\cE{\mathcal{E}}

\def\I{\bm{1}}

\newcommand\bel[1]{\begin{equation} \label{#1}}
\newcommand\ee{\end{equation}}
\newcommand\bea{\begin{eqnarray*}}
\newcommand\eea{\end{eqnarray*}}
\newcommand\er{\eqref}
\newcommand{\nn}{\nonumber}

\newcommand\su{\mathfrak{su}}
\newcommand\gl{\mathfrak{gl}}
\newcommand\h{\mathfrak{h}}

\newcommand{\der}{{\mathfrak {der}}}

\renewcommand\Im{\operatorname{Im}}
\newcommand\tr{\operatorname{tr}}

\newcommand\ad{\mathrm{ad}}
\newcommand\rank{\operatorname{rank}}
\newcommand\ch{\operatorname{ch}}

\newtheorem{definition}{Definition}
\newtheorem{theorem}{Theorem}

\newtheorem{conjecture}{Conjecture}
\theoremstyle{remark}
\newtheorem{remark}{Remark}

\newcommand\Superop{\cE}
\newcommand{\mtil}{\tilde{m}}

\begin{document}

\title{Quantum Channels and Representation Theory}

\author{William Gordon \surname{Ritter}}
\affiliation{Harvard University Department of Physics \\
    17 Oxford St., Cambridge, MA 02138}

\date{February 9, 2005}

\begin{abstract}
In the study of $d$-dimensional quantum channels $(d \geq 2)$, an
assumption which is not very restrictive, and which has a natural
physical interpretation, is that the corresponding Kraus operators
form a representation of a Lie algebra. Physically, this is a
symmetry algebra for the interaction Hamiltonian. This paper
begins a systematic study of channels defined by representations;
the famous Werner-Holevo channel is one element of this infinite
class. We show that the channel derived from the defining
representation of $SU(n)$ is a depolarizing channel for all $n$,
but for most other representations this is not the case. Since the
Bloch sphere is not appropriate here, we develop technology which
is a generalization of Bloch's technique. Our method works by
representing the density matrix as a polynomial in symmetrized
products of Lie algebra generators, with coefficients that are
symmetric tensors. Using these tensor methods we prove eleven
theorems, derive many explicit formulas and show other interesting
properties of quantum channels in various dimensions, with various
Lie symmetry algebras. We also derive numerical estimates on the
size of a generalized ``Bloch sphere'' for certain channels. There
remain many open questions which are indicated at various points
through the paper.
\end{abstract}



\maketitle

\tableofcontents

\section{Introduction}

It has been known for more than three decades through the work of
\citet{Kraus}, and that of \citet{Choi}, that a general channel
admits an operator-sum representation, so its action on an
$n\times n$ density matrix $\rho$ has the form $\sum_\mu M_\mu
\rho M_\mu^\dag$. A special case occurs when the $M_\m$ are
Hermitian and maximal in number, so that $\mu = 0\ldots$
\mbox{$n^2-1$}.

Define $\la_0 = \I$ and let $\la_a$ for $a \geq 1$
denote the $n$-dimensional Gell-mann matrices, which are standard
generators for the Lie algebra $\su_n$.
Then $\{ \la_\m  \}$ is a basis for the
space of Hermitian matrices. Taking $M_0$ proportional to the
identity, there exist constants $U_{ab}$ such that
\[
    M_a = \sum_b U_{ab} \la_b.
\]
If $U$ is a unitary matrix, then we may take each $M_\m$
proportional to $\la_\m$ without changing the quantum channel
defined by these Kraus operators.

In the latter case, one may readily calculate $\sum_\mu M_\mu \rho
M_\mu^\dag$, since $\rho$ itself may be expanded in the $\la_\m$
basis, and the Gell-mann matrices in any dimension satisfy elegant
product identities. Indeed, we calculate this operator explicitly
in Sec.~\ref{sec:su_n_channel}. Exploiting the product identities leads
to a simple, elegant and explicit formula for the action of the
$\su_n$ channel; if $\rho = \f{1}{d} \I + \sum_{\m} v_\m \la_\m$
then the channel multiplies $v$ by a scalar. Thus the
assumption that the Kraus operators
are proportional to generators of $\su_n$ is extremely useful as
a calculational tool, but it is not merely that. A quantum channel
models the interaction of a decohering system with its environment,
and the identification of the $M_\mu$ as generators of a Lie algebra
is related to a symmetry of the interaction Hamiltonian. See
\citep{Lidar:1998hs} and \citep{LW} for details.

The generalizations of the calculations in $\su_n$ described above
to other Lie algebras and to higher-dimensional representations
are illuminating, and have not appeared in the literature before.
These cases necessarily have the property that not all of the
Hermitian matrices in that dimension are linear combinations of
representation matrices, so direct generalization of the
calculational method outlined above for $\su_n$ will not work, and
a new idea is required. This is the subject of Section
\ref{sec:other-reps}. We give a particularly detailed analysis of
the three-dimensional representation of $\su_2$; however, many of
the formulas we use there generalize readily to higher spin. In an
interesting twist, the spin-one case turns out to be a
generalization of the Werner-Holevo channel, and suggests that Lie
algebra channels may play an interesting role as examples or
counterexamples to the well-known AHW conjecture.

Section \ref{sec:g2} analyzes the channel based on the exceptional
algebra $\g_2$. In Section \ref{sec:blochmanifold} we generalize
some aspects of the Bloch sphere to density matrices constructed
from Lie algebra representations. It is shown that for each
representation, there is a class of density matrices parameterized
by a closed, bounded (hence compact) submanifold of Euclidean space,
which we term `the Bloch manifold.' Explicit bounds are given on the size of
these manifolds. A general method is given for finding the Bloch manifold
exactly, using trace identities.

\section{Kraus Decomposition}

A quantum channel is a model for a particular snapshot of the time
evolution of a density matrix, and especially for the evolution of
pure into mixed states. Let $\cH$ be a Hilbert space of dimension
$n$, and let $\gl(\cH)$ denote the vector space of all linear
transformations from $\cH \to \cH$. A map $\Superop : \gl(\cH) \to
\gl(\cH)$ is said to be \emph{completely positive (CP)} if it is linear
and $\Superop \otimes \I$ is positive on $\cH \otimes \cH'$ for
all $\cH'$. The map is said to be \emph{trace-preserving} if
$\Tr\, \Superop(A) = \Tr\, A $ for all $A \in \gl(\cH)$.

\begin{definition} \label{def:superop}
A \emph{CPT map} or \emph{stochastic map} is a completely
positive, trace-preserving linear transformation $\Superop :
\gl(\cH) \to \gl(\cH)$.
\end{definition}

Of central importance to the current work is a famous theorem of
\citet{Kraus} which gives a simple representation of any quantum
channel.

\begin{theorem}[Kraus Decomposition Theorem] \label{thm:kraus}
For any quantum channel $\Superop$, there exists a finite set of
operators
\[
    M_0, M_1, M_2, \ldots, M_k, \ \te{ where } \ k \leq (\dim \cH)^2,
\]
such that
\bel{kraus}
    \Superop(\brho)
    =
    \sum_\mu M_\mu \brho M_\mu^\dag
    \quad
    \te{ with }
    \quad
    \sum_\mu M_\mu^\dag M_\mu  = \I .
\ee
In this situation, \er{kraus} is called the \emph{Kraus
representation}, the \emph{operator sum representation}
or the \emph{Stinespring form}, while
$\sum_\mu M_\mu^\dag M_\mu  = \I$ is sometimes called the
\emph{normalization condition} and is just the statement that the
map is trace-preserving.
\end{theorem}

A proof of this theorem may be found in the original article of
\citet{Kraus}, or in the book by \citet{NielsenChuang}. We simply
note that the converse, namely that any operator of the form
\er{kraus} satisfies the conditions of Definition
\ref{def:superop}, is clearly true. A stochastic map may also be
obtained as the partial trace of a unitary conjugation on a larger
space; see \citep[Sec.~III.D]{Ruskai} for a discussion. The
representation \er{kraus} is sometimes called the \emph{Stinespring
form} since its existence follows from the Stinespring dilation theorem
\citep{Stinespring}.

This is a general framework, and in order to obtain explicit
results, further assumptions are necessary. A mathematically
elegant assumption is that the possible errors introduced in the
decoherence process are not arbitrary, but that they correspond to
the action of the infinitesimal generators of a Lie group $G$ of
continuous symmetries. This provides a simple model for symmetry
breaking in quantum mechanics.

The situation just described, in which the error generators are
also generators for a matrix representation of a semisimple Lie
algebra, follows naturally from the model of Markovian dynamics
considered by \citet*{Lidar:1998hs}. This was shown to have
important consequences for the possibility of decoherence-free
dynamics; see \citep[and references therein]{LW} for an up-to-date
review. The present work may be considered as a further
exploration of the consequences of that model, for a snapshot of
the time evolution.

The \emph{qubit depolarizing channel} is a model of a decohering
qubit in which the decoherence has an $SU(2)$ symmetry. With
probability $1-p$ the qubit remains intact, while with probability
$p$ an error occurs. The error can be one of three types, each
equally likely. These errors are implemented by applying Pauli
matrices to the qubit state. In other words, an error involves
applying one of the generators of the Lie algebra $\su_2$ to a
vector in its irreducible two-dimensional representation. These
generators admit direct physical interpretations as bit-flip
errors, phase-flip errors, or combinations of those.

The qubit depolarizing channel admits a generalization to a
channel with $k$ possible errors based on an $n$-dimensional
representation $\cH$ of a $k$-dimensional Lie algebra $\g$, for
which detailed properties have not been previously investigated,
and which is the main topic of the present work.
As we develop the general theory of these channels in the
following sections, we will see that not all Lie algebras can give
quantum channels (at least not in the way outlined here),
and even for Lie algebras which do give
channels, not all representations are acceptable. For $\g$
semisimple, it is necessary that the quadratic Casimir take a
single value on all elements of the representation space. This
holds for all irreducible representations, and some reducible
ones. The non-semisimple case is more difficult, and its treatment
will be deferred to a separate paper.

\section{Quantum Channels From Lie Algebra Representations}
\label{sec:basics}

This section contains our notations and conventions for the
generalized depolarizing channels which will be studied in detail
in later sections. The possibility of defining a quantum channel
based on a representation of a compact Lie algebra was mentioned
briefly, but never elaborated upon, in a paper of \citet{GW2003}.
In any case, it is not necessary that the Lie algebra be compact.

\subsection{Pure Lie Algebra Channels}
\label{sec:pure}

It is a standard convention \cite{MacFarlane,Georgi} to normalize
the canonical generators for the defining representation of
$\su_n$ so that
\bel{2delta}
    \Tr\lrp{ \la_a \la_b } = 2 \delta_{ab} .
\ee
This has the desirable feature that the canonical generators for
$n=2$ are the Pauli matrices, and those for $n=3$ are the familiar
Gell-mann matrices, while inserting factors of 2 in certain
formulae. With convention \er{2delta}, these generators will be
orthogonal but not orthonormal with respect to the Killing form.
We return to this point below.

On a general semisimple Lie algebra, the Killing form $K$ is
defined as
\[
    K(X,Y) = \Tr( \ad(X) \circ \ad(Y) : \g \to \g ) ,
\]
where the trace is taken in the adjoint representation. At the
moment we focus on semisimple algebras $\g$, for which the Killing
form is nondegenerate, and return to treat non-semisimple algebras
in a later section. Let $\alpha$ be an irreducible representation
of $\g$, let $X_i$ be any basis of $\g$, and let $X_i'$ denote the
dual basis with respect to the Killing form. The Casimir operator
\[
    C_2(\alpha) = \sum_i \alpha(X_i) \alpha(X_i')
\]
does not depend on the choice of basis, and by Schur's lemma is
proportional to the identity, so we write \mbox{$C_2(\alpha) =
c_2(\alpha) \I$}. If $X_i$ is orthonormal with respect to $K$,
then $C_2(\alpha) = \sum_i \alpha(X_i)^2$. For reducible
representations, $C_2(\alpha)$ may not be proportional to the
identity.

\begin{definition} \label{def:gdc}
Let $\g$ denote a Lie algebra of dimension $k$, with basis $\{ X_i
: i = 1, \ldots, k \}$. Let $\alpha$ be an irreducible
$\g$-representation on the Hilbert space $\cH$. The
\emph{generalized depolarizing channel} or \emph{Lie algebra
channel} is defined to be the channel in which an error occurs
conditionally with probability $p$, causing an initial state
$\ket{\psi} \in \cH$ to evolve into an ensemble of the $k$ states
$\alpha(X_i) \ket{\psi}$, all with equal likelihood.
\end{definition}

The Kraus operators for the channel of Definition \ref{def:gdc}
are given by
\bel{theMs}
    M_0 = \sqrt{1-p} \ \I, \quad
    M_i =  \sqrt{\La p}\ \alpha(X_i) \, .
\ee
where $\La$ is a normalization constant which will be fixed
momentarily. The operators $M_\mu$ are hermitian if the
representation is unitary and if $p \in [0,1]$, and are
constrained to satisfy $\sum_\mu M_\mu M_\mu = \I$, which fixes
the value of the constant $\La$ appearing in \er{theMs}.
By definition,
\[
    \sum_\mu M_\m^2 =
    (1-p) \I + \La p \sum_i \alpha(X_i)^2 .
\]
If $\sum_i \alpha(X_i)^2 = Z\cdot \I$, where $Z$ is a constant
(which in most cases we can take to be real), then
\[
    \La = \f{1}{Z} \ .
\]
If $X_i$ is orthonormal with respect to the Killing form, then $Z
= c_2(\al)$. More generally, if the basis satisfies
\[
    K(X_i, X_j) = \fN\, \delta_{ij} \, , \quad
    \fN > 0,
\]
then it can be rescaled to an orthonormal basis by a single
constant. In this situation,
\bel{La}
    Z = \fN \, c_2(\al),
    \quad
    \La = \f{1}{Z} \, .
\ee
Defining the Killing norm by $\nrm{x}_K^2 = K(x,x)$, we note that
if
\[
    \nrm{X_i}_K \ne \nrm{X_j}_K,
\]
for some pair of indices $i,j$, then the normalization condition
cannot be satisfied.

What if the representation is reducible? Suppose $\cH = V \oplus
W$ as a direct sum of irreducible $\g$-modules, and $X_i$ is
orthonormal with respect to $K$. Then there exist independent
constants $Z_V$ and $Z_W$ such that the operator
\[
    C_2(\al) = \sum_i \al(X_i)^2 =
    \begin{pmatrix}
    Z_V & 0 \\ 0 & Z_W
    \end{pmatrix}
\]
as a block decomposition on $V \oplus W$. If \mbox{$Z_V \ne Z_W$},
then it is not possible for the Kraus operators \er{theMs} to give
a trace-preserving map. On the other hand, if $Z_V = Z_W$ then
they do define a CPT map even though the representation is
reducible.

What if $p > 1$? Then $M_0 = i \displaystyle\sqrt{p-1} \ \I$, and
we have
\[
    \sum_\mu M_\mu M_\mu^\dag
    =
    (2p-1) \I \, .
\]
Thus the map cannot be trace-preserving unless $p = 1$, which is a
contradiction. A similar argument shows that $p < 0$ does not give
a trace-preserving map. Thus, if we wish to study the framework of
Definition \ref{def:gdc}, then we must limit ourselves to $p \in [0,1]$.

We summarize the results of the last few paragraphs in a Theorem.

\begin{theorem}[Normalization]
\label{thm:normalization}
Consider the Kraus operators
\[
    M_0 = (1-p)^{1/2} \I,
    \ \te{ and } \
    M_i =  (\La p)^{1/2} \alpha(X_i),
\]
for $i = 1 \ldots k$. If
\begin{enumerate}
\item[(i)] $p \in [0,1]$,
\item[(ii)] The representation $\al$ of $\g$ is a direct sum of
irreducible representations all with the same quadratic Casimir,
and
\item[(iii)]
$\exists\ \fN > 0$ such that $K(X_i, X_j) = \fN\, \delta_{ij}$ for
all $i,j$,
\end{enumerate}
then $\sum_\m M_\m M_\m = \I$ with $\La$ given by eqn.~\er{La}.
Conversely, if any of (i)-(iii) is not satisfied, then (except in
trivial cases) there does not exist $\La$ s.t. $\sum_\m M_\m M_\m
= \I$, and the $M$'s do not give rise to a quantum channel.
\end{theorem}

The coefficients of the $M_\mu$ in \er{theMs} admit a natural
``probability of error'' interpretation, but in Section
\ref{sec:extensions} we investigate the possibility of modifying
them to complex coefficients in order to obtain a new channel. We
find that no new channels arise unless one is willing to promote
the coefficients to operators.

Using \er{kraus}, the Lie algebra channel has the explicit Kraus
decomposition
\bel{superop}
    \brho \to \Superop(\brho) = (1-p) \brho + \f{p}{Z}
    \sum_{i=1}^k \al(X_i) \,\brho\, \al(X_i)
    \, .
\ee
As is proven in standard textbooks \citep[see Theorem
8.9]{Georgi}, the trace of any generator of any representation of
a compact simple Lie algebra is zero, so in particular, the
$\al(X_i)$ are traceless. Moreover, it is clear that this
transformation satisfies the defining properties for a quantum
channel, given here as Definition \ref{def:superop}.

Two operator-sum representations
\[
    \sum_\mu M_\mu \brho M_\mu^\dag
    \quad
    \te{and} \quad
    \sum_\nu N_\nu \brho N_\nu^\dag
\]
describe the same channel if and only if there exists a unitary
matrix $U_{\nu\m}$ such that $N_\nu = U_{\nu\m} M_\m$. Therefore,
it is immaterial which basis of the Lie algebra that we use, as
long as the two bases are related by a $U(N)$ similarity
transformation. As noted in Theorem \ref{thm:normalization}, in
order to build a channel satisfying the normalization condition,
we are forced to use a basis satisfying ``orthonormality,''
$K(X_i,X_j) = \fN \,\de_{ij}$. But any two ``orthonormal'' bases
in this sense are related by a unitary transformation, so the
CPT map constructed above is independent of the basis chosen
for $\g$.

Given a Lie algebra $\g$ and a representation $\alpha$ on a vector
space of dimension $d$, the CPT map \er{superop} is a model
for decoherence through a $d$-level noisy quantum channel, with
errors that are not completely arbitrary; rather, they transform
the state in a way determined by the representation of $\g$.

The channels \er{superop} have an extremely interesting structure.
For a certain subclass of possible Lie algebra representations,
the channel \er{superop} has an action which, like the qubit case,
is most simply described by a Bloch parameterization with
polarization vector $v \in \R^k$, where $k = \dim \g$. In these
cases, we show that \er{superop} decreases the length of $v$, and
so deserves the title `generalized depolarizing channel.' In other
cases of interest, a single Bloch vector is not sufficient, but
the action of the channel can be described by similar rescalings
of symmetric 2-tensors or higher-rank objects.

A natural step, which we begin in the next section, is to
calculate the expression \er{superop} explicitly in certain
representations of classical Lie algebras.

\begin{remark}
When we use the terminology ``the \mbox{$\g$-channel},'' where
$\g$ is a semisimple Lie algebra, the fundamental representation
of $\g$ is implied. Examples of fundamental representations
include the $n$-dimensional defining representation of $\su_n$,
and the 7-dimensional irrep of $G_2$.
\end{remark}

It is easy to see that the Lie algebra channel \er{superop} always
has the property of being \emph{doubly stochastic}, i.e.
\mbox{$\Superop(\I) = \I$}. See for example \citep{GW2003} for further
discussion.

\subsection{A Note on Coefficients and Extensions}
\label{sec:extensions}

As discussed prior to Theorem \ref{thm:normalization}, for $p
\not\in [0,1]$ the channel defined by \er{superop} is CP but not
T, and it is possible to recover a CP channel only if we consider
different coefficients for the Kraus operators \er{theMs}. To this
end, let us first consider
\bel{general-M1}
    M_0 = m_0 \I \ \ \te{ and } \ \
    M_i = \f{\mtil}{\sqrt{Z}}\, \alpha(X_i) \, ,
\ee
where $m_0, \mtil \in \C$ are some complex constants. Then to
obtain a trace-preserving map, we require
\[
    \sum_\m M_\m^\dag M_\m
    =
    |m_0|^2 + |\mtil|^2
    = 1 \, .
\]
This condition is equivalent to the statement that the point
$(m_0,\mtil) \in \C^2 \cong \R^4$ lies in the unit 3-sphere $S^3
\subset \R^4$.

We can now view the coefficients of the Kraus operators \er{theMs}
as the projection $S^3 \to S^1$. Introduce a parameter $q \in
[-1,1]$ such that $p = q^2$, and write \er{theMs} as $M_0 = \pm
{\displaystyle\sqrt{1-q^2}}\ \I,$ and $M_i = (q/\sqrt{Z})
\alpha(X_i)$. Then ignoring $\sqrt{Z}$, the coefficients of $M_i$
and $M_0$ give a point on the unit circle. Further, $m_0$ and
$\mtil$ only enter through the square of their magnitude, so the
two additional parameters associated to projecting from the
3-sphere are fictitious, and \er{superop} is in fact the most
general channel of this kind.

A non-trivial generalization is obtained by promoting $m_0$ and
$\mtil$ to \emph{operators}. However this ``generalization'' is a
special case of a well-known operation which extends an existing
channel $\cE_B$ using any set of operators which satisfy the
normalization condition \er{kraus}. Given two sets of Kraus
operators $A_1, \ldots, A_r$ and $B_1, \ldots, B_s$ acting on the
same vector space and satisfying
\[
    \sum_{i=1}^r A_i^\dag A_i
    =
    \sum_{j=1}^s B_j^\dag B_j
    =
    \I\, ,
\]
we note that the set of operators
\bel{extended-channel}
    \{ A_1, \ldots, A_{r-1}, B_1 A_r, \ldots, B_s A_r \}
\ee
also satisfies the normalization condition, because
\[
    \sum_{j=1}^s (B_j A_r)^\dag B_j A_r
    =
    A_r^\dag A_r \, .
\]
This construction is \emph{natural} with respect to the channel
$\Superop_B$ defined by $B_i$, in the sense that if $\{B_i'\}$ is
another set of Kraus operators defining the same channel, then the
channel defined by \er{extended-channel} is also the same.
Naturality does not hold for the $A$ operators, but this will not
concern us here. We call this procedure \emph{the extension of
$\Superop_B$ by the A-operators, on the element $A_r$}.

For example, one may notice that the operators $Z^{-1/2}
\alpha(X_i)$ of the previous section satisfy the normalization
condition since the sum of their squares is a Casimir element, and
the normalization constant $Z$ cancels the numerical factor.
Consider this the B-channel, and extend it on every element by the
same set of Kraus operators. This yields a ``double $\g$-channel''
with Kraus operators
\bel{tens-ext}
    \Big\{ \f{1}{Z} X_i X_j \ :\  i,j = 1\ldots k \Big\} \, .
\ee
These operators generate the image of $\g \otimes \g$ under the universal
homomorphism expressed in the commutative diagram \er{commdiag}.

This underscores the fact that, aside from the basic examples of
new quantum channels provided by Section \ref{sec:pure}, many
further examples may be obtained by extension, as in
\er{extended-channel}. As in the basic Lie algebra channels,
computations with extended channels are facilitated by the
existence of non-trivial identities which exist among the
representation matrices. Channel \er{tens-ext} is interesting
because for many representations, the matrices $\alpha(X_i)$ do
not span the entire space of traceless $d \times d$ matrices, but
the set of products $\alpha(X_i) \alpha(X_j)$ spans a subspace of
larger dimension. Therefore the extension leading to \er{tens-ext}
is a way of generating a channel whose Kraus operators come closer
to spanning the space of all matrices in the appropriate
dimension. If a density matrix were written as $\rho = \sum_{ij}
w_{ij} X_i X_j$, and if the representation satisfies an identity
for reduction of products of six generators, then one can
calculate the action of \er{tens-ext} on $\rho$ explicitly.

We are now in a position to interpret the channel defined by
\er{general-M1} with complex coefficients $m_0, \mtil$ as the
extension \er{extended-channel}  of the nontrivial Lie algebra
channel $B_i = Z^{-1/2} \alpha(X_i)$ by the identity channel with
the unusual Kraus representation $A_1 = m_0 \I$, $A_2 = \mtil \I$.
In case $\mtil = q \in [-1,1]$ and $m_0 = \pm \sqrt{1-q^2}$ we
recover \er{superop}. Given any channel whose set of Kraus
operators do not contain $\I$, we can always extend it so that
they do contain the identity by this method.

For the rest of this paper, we will assume that the Kraus operators
take the form \er{theMs} in order to retain the beautiful
probabilistic interpretation given by Definition \ref{def:gdc}.
As we continue, we will keep the fact in mind that extensions
are possible, and develop methods which easily generalize.

\section{The SU${}_{\bm{n}}$ Channel} \label{sec:su_n_channel}

The $\su_n$ channel, our first example, is the channel built from
the $n$-dimensional defining representation (also called `standard
representation') of $\su_n$. It is simpler than most other
channels studied in this paper, because it admits a complete
solution. Its action on any arbitrary input density matrix can be
calculated in closed form using the Bloch parameterization, and in
all cases it is a depolarizing channel.

One reason for the beauty and simplicity of the $\su_n$ channel is
that any $n$-dimensional density matrix admits a Bloch vector
parameterization in terms of $\su_n$ generators. This is because
$k \equiv \dim(\su_n) = n^2 - 1$ is only one less than $n^2$, the
dimension over $\R$ of the space of $n \times n$ Hermitian
matrices.

Any $n \times n$ Hermitian matrix $\rho$ may be represented as
\[
    \rho = \f{1}{n} \lrp{ \tr(\rho) \I + T},
    \qquad T \in \su_n \, ,
\]
and having chosen a basis $X_a$ for $\su_n$, it follows that
\[
    T = \sum_{a=1}^{k} v_a X_a \equiv v \cdot X\, ,
\]
for some coefficient vector $v$. In analogy with the well-known
parameterization of the $2 \times 2$ density matrices as the
interior of a sphere, we will refer to $v$ as the \emph{Bloch
vector}.

For $n \geq 3$ it may be hard to visualize the geometry of the
space of density matrices in terms of the geometry of $v$. This
question was first considered in the $n=3$ case by
\citet{MacFarlane}. Section \ref{sec:blochmanifold} undertakes a
systematic study of the geometry of the space of $v$ which lead to
a valid density matrix in various representations. We call this
space the \emph{Bloch manifold} and give details of the geometry
for a number of important examples, including all representations
of $\su_2$, and the $n$-dimensional irrep of $\su_n$.

In this section, we take $\alpha$ to be the standard
representation of $\su_n$ on a vector space $\cH$ of dimension
$n$. For simplicity, we let $X_i$ denote both the generator of
$\su_n$ and its image under this representation. One could now
compute the quadratic Casimir in the standard way using roots and
weights, but it will turn out that the value of this Casimir as
well as all other properties we will need to obtain a complete
solution to the $\su_n$ channel follow from the single relation
\bel{beta}
    X_i X_j = \beta \delta_{ij} \I + \sum_k Q_{ijk} X_k \, .
\ee
for some constant $\beta$ and tensor $Q_{ijk}$. Of course, this
relation is just the decomposition of a Hermitian matrix into a
trace part with trace $n \beta \de_{ij}$, and a linear combination
of the $X_k$, which generate the space of traceless matrices.

Elements of the standard basis of $\su_n$ are called Gell-mann
matrices, and they satisfy
\[
    \Tr(X_i X_j) = 2 \de_{ij},
\]
so $\beta = 2/n$. Many properties of the $Q$ tensor already follow
from the single assumption that $X_i$ generate a Lie algebra. It
is immediate that $Q_{[ij]k} = i f_{ijk}$ where $[ij]$ denotes
antisymmetrization, and $f_{ijk}$ is 1/2 times the structural
tensor of the Lie algebra. It follows that
\[
    Q_{ijk} = d_{ijk} + i f_{ijk}
\]
for some $d_{ijk}$ symmetric in the first two indices. Also,
\er{beta} implies
\[
    \{ X_i, X_j \} = \f{4}{n} \de_{ij} \I + 2 \sum_l d_{ijl} X_l .
\]
Multiplying by $X_k$ and taking the trace yields
\[
    d_{ijk} = \f14 \Tr(\{ X_i, X_j \} X_k)  ,
\]
therefore the $d$-tensor is \emph{completely} symmetric, and
interchange of any two indices has the effect of complex
conjugating $Q$. Since $\sum_i X_i X_i$ is a multiple of the
identity,
\bel{dnotrace}
    \sum_i d_{iik} = \f12 \Tr\Big( (\sum_i X_i X_i)  X_k\Big) = 0
\ee

It follows from the associativity of matrix multiplication that
\[
    f_{ijm} f_{klm} =
    \f{2}{n} (\de_{ik} \de_{jl} - \de_{il} \de_{jk})
    + d_{ikm} d_{jlm} - d_{jkm} d_{ilm} \ .
\]
with a sum over $m$ implied. Contracting $j$ and $k$ and using
\er{dnotrace} yields
\[
    d_{ijm} d_{ljm} = f_{ijm} f_{jlm} + (2n-\f{4}{n}) \de_{il}
     \ ,
\]
By a general property of compact semi-simple Lie algebras, the
structure constants satisfy
\bel{fprod}
    f_{ijk} f_{ljk} = n \de_{il} \, ,
\ee
Therefore, \mbox{$d_{ijm} d_{ljm}$} $ = $
\mbox{$\displaystyle(n-\f{4}{n}) \de_{il}$}. Using this and
\er{fprod}, we obtain
\bel{QQ}
    Q_{ijm} Q_{ljm} = d_{ijm} d_{ljm} - f_{ijm} f_{ljm}
    =
    - \f{4}{n} \de_{il}\, .
\ee

For this basis of $\su_n$, $Z = 2k/n$, where $k = n^2-1$. The
action of the channel
\[
    \brho \to \Superop(\brho)
    =
    (1-p) \brho + \f{pn}{2k} \sum_{i=1}^k X_i \brho X_i
\]
on the density matrix
\[
    \brho = \f{1}{n} \lrp{\tr(\brho) \I + v \cdot X}
\]
is given by
\bel{triple}
    \Superop(\brho) =
    \f{\tr(\brho)}{n} \I + \f{1-p}{n} v \cdot X
    +
    \f{p}{2k} \sum_{i,j} v_j X_i X_j X_i
    \, .
\ee
Using \er{beta} to expand the triple product, we have
\[
    \sum_{j,i} v_j X_i X_j X_i
    =
    \beta v \cdot X + \beta \sum_{i,j} v_a Q_{iji} \I
    + \sum_{i,j,k,a} v_j Q_{ijk} Q_{kia} X_a \, .
\]
Since $\Superop(\brho)$ has unit trace, it must be the case that
\mbox{$\sum_i Q_{iji} = 0$}. The same conclusion also follows from
\er{dnotrace}, but it is amusing to see that $\sum_i Q_{iji}$ must
vanish because this is a CPT map. Therefore,
\bel{dep-chan1}
    \Superop(\brho)
    =
    \f{\tr(\brho)}{n} \I + \f{1-p+p/k}{n} v \cdot X
    +
    \f{p}{2k} \sum_{i,j,k,a} v_j Q_{ijk} Q_{kia} X_a
    .
\ee
Using \er{QQ}, we have finally
\[
    \Superop(\brho_v)
    =
    \f1{n} \lrp{\tr(\brho) \I + f(p,n)\, v \cdot X} ,
\]
where
\bel{fpn}
    f(p,n) = 1 - p - \f{p}{k} =  \f{(1-p) n^2-1}{n^2-1} \,.
\ee
In the qubit case, $f(p,2) = 1-4p/3$, which is consistent with
standard results.

The $\su_n$ channel maps an initial density matrix to a linear
combination of itself and the identity, i.e. it has the form
\bel{lambda}
    \De_\la(\brho) = \la \brho + \lrpBig{ \f{1-\la}{n} } \I \, .
\ee
This is the standard definition of the $n$-dimensional
depolarizing channel. The information-carrying capacity of this
channel was studied in great detail by \citet{King}, where notably
the Amosov-Holevo-Werner conjecture was established for channels
which are products of a depolarizing channel with an arbitrary
channel. Channels based on representations of semisimple algebras
generically do not take the form \er{lambda}, except possibly on
special subsets of the space of density matrices. See Section
\ref{sec:other-reps} and in particular Theorem \ref{thm:spin1} for
a Lie algebra channel that is \emph{not} a depolarizing channel.

The depolarizing channel on an $n$-dimensional Hilbert space
satisfies complete positivity if and only if
\[
    \f{1}{1-n^2} \leq \la \leq 1 \, .
\]
The $\su_n$ channel has the form \er{lambda} for $\la = f(p,n)$.
Note that the relation
\[
    \f{-1}{n^2-1} \leq f(p,n) \leq 1
\]
holds for all $n \geq 2$. In fact, $f(p,n)$ saturates both of
these inequalities at the endpoints of the allowed range, $0 \leq
p \leq 1$.

At the special value $p = 1 - n^{-2}$, the $\su_n$ channel is a
constant map from $\R^{n^2-1}$ into the space of density matrices:
\bel{su_n-critical}
    \Superop(\brho_v) =
    \f1{n} \I \ \te{ for all } v,
    \te{ at }
    p = p_c \equiv 1 - n^{-2} .
\ee
Physically, if the probability of error happens to be $p = p_c$,
then $\su_n$-decoherence evolves an arbitrary initial density
matrix into a completely uniform ensemble consisting of pure
states with equal probabilities. This is the ``worst'' value of
$p$, in the sense that all information about the initial density
matrix has been lost. This result is \emph{stable} in the sense
that if $p$ is only approximately equal to the critical value, the
initial density matrix decoheres into an approximately uniform
ensemble.

We will see in Sec.~\ref{sec:other-reps} that for other Lie
algebra channels, there are multiple critical values of $p$ which
generalize \er{su_n-critical}; this analysis leads to an
interesting decomposition of the space of density matrices on
$\cH$ which is discussed in Section \ref{sec:critical}.

\section{Other Representations} \label{sec:other-reps}

\subsection{General Remarks}

In the $n$-dimensional standard representation of $\su_n$, the
representation matrices $\al(X_i)$ span the space of all traceless
Hermitian matrices, and thus an arbitrary initial density matrix
can be expressed in terms of the $\al(X_i)$ and the identity. As
we consider higher-dimensional representations, the representation
matrices become increasingly sparse in the space of all traceless
matrices, and thus only some fraction of the set of all possible
density matrices can be expressed in the form $d_\al^{-1} \I +
\sum_i v_i \al(X_i)$. This is not all the bad news; for
higher-dimensional irreducible representations (irreps), there is
generally no analogue of the identity \er{beta} which holds for
$\su_n$.

Therefore, the simple calculations we have done for the
$n$-dimensional irrep of $\su_n$ do not generalize in any simple
way to other representations; new ideas are needed. In this
section, we develop methods for dealing with the general case of
arbitrary representations. Let $d = d_\al$ denote the dimension of
the representation $\al$, and $\gl_d$ as usual denotes the
associative algebra of all $d \times d$ matrices.

A representation $\phi$ of $\g$ lifts to a unique associative
algebra homomorphism $\tilde{\phi}$ of the universal enveloping
algebra $\cU(\g)$, by the universal property most elegantly
expressed in the commutative diagram
\bel{commdiag}
    \xymatrix{ \g \ar^i[r] \ar_{\phi}[dr] & \cU(\g) \ar^{\tilde{\phi}}[d] \\
    \ & \gl_d }
\ee
The action of $\tilde{\phi}$ is simply to convert the tensor
product to matrix multiplication, i.e. \mbox{$\tilde{\phi}(x
\otimes y)$} $=$ \mbox{$\phi(x) \cdot \phi(y)$}, etc. The
interesting property about this commutative diagram, and one which
gives a computational method for Lie algebra channels, is that if
$\phi$ is an irreducible faithful representation and if $\g$ is a
semisimple Lie algebra, then $\tilde{\phi}$ is surjective.

This surjectivity has the consequence that for \emph{any
representation} of said Lie algebra, \emph{every} density matrix
can be represented as a linear combination of products of the
representation matrices. In other words, the new calculational
method outlined in this section will always work. Before
continuing our discussion of this, let us consider a simple but
nontrivial example, the spin-1 channel, in complete detail.

\subsection{The Spin-1 Channel}

Consider the spin-1 representation of $\su_2$. We use standard
angular momentum notation, in which
\[
    J_1 = \f{1}{\sqrt{2}}
    \threemat{ 0 & 1 & 0 \\ 1 & 0 & 1 \\ 0 & 1 & 0 },
    J_2 = \f{1}{\sqrt{2}}
    \threemat{ 0 & -i & 0 \\ i & 0 & -i \\ 0 & i & 0 },
    J_3 =
    \threemat{ 1 & 0 & 0 \\ 0 & 0 & 0 \\ 0 & 0 & -1 }.
\]
Before generalizing to arbitrary density matrices, we restrict
attention to the simpler example of density matrices $\brho$ which
are of the form
\bel{Blochrho}
    \brho_v = \f{1}{3} \lrp{\I + v \cdot J}, \quad
    v \in \R^3 .
\ee
Then
\bel{spin1a}
    \Superop(\brho_v) = \f13 \I + \f{1-p}{3} v \cdot J +
    \f{p}6 \sum_{a,b} v_b J_a J_b J_a
    \, .
\ee
The relation analogous to \er{beta} does not hold, i.e. $J_a J_b$
is not a linear combination of $\I$ and $\{ J_i : i = 1 \ldots 3
\}$. In this special case, the triple product appearing in
\er{spin1a} simplifies considerably;
\bel{tripleproduct}
    J_a J_b J_a = \delta_{ab} J_a \ \te{(no sum)},
\ee
which implies that
\bel{spin1b}
    \Superop(\brho_v) = \f13 \lrp{ \I + \big(1 - \f{p}2\big) v \cdot J}
    \, .
\ee
This takes the form \er{Blochrho} with $v \to \big(1 - \f{p}2\big)
v$. Thus, for \mbox{$3 \times 3$} density matrices admitting a
Bloch parameterization, \emph{if $p$ is a probability} then the
spin-1 channel scales the Bloch vector by a number between 1/2 and
1.

Interestingly, we can go further and find a Bloch-type picture of
the spin-1 channel on a general density matrix. The six elements
of the form
\[
    J_{(a} J_{b)} \equiv \f12 \lrp{ J_a J_b + J_b J_a } \, ,
\]
together with $J_1, J_2,$ and $J_3$, span the space of $3\times 3$
matrices. Therefore an arbitrary \mbox{$3 \times 3$} density
matrix $\brho$ can be written as
\bel{rhovw}
    \brho_{v,w} = v\cdot J + \sum_{a,b} w_{ab} J_{(a} J_{b)}
\ee
for some vector $v$ and symmetric tensor $w$.

We use standard physics normalizations which entail that for the
spin $s$ representation in $d = 2s+1$ dimensional space,
\[
    \sum_a J_a^2 = \la\, \I
    \quad
    \Rightarrow
    \quad
    \tr( J_a J_b ) = \f{d \la}{3} \ \de_{ab} .
\]
where $\la = s(s+1)$. Then we have
\[
    \tr(\brho_{v,w}) = \f{d \la}{3} \tr(w).
\]
It follows that in order to have a density matrix, we require
$\tr(w) = 3(d \la)^{-1}$. For $s = 1$, $\tr(w) = 1/2$.

\begin{theorem}[Action of the Spin-1 Channel] \label{thm:spin1}
The action of the spin-one channel on the vector and symmetric
tensor are $v \to v'$ and $w \to w'$, where
\begin{eqnarray}
    v_a &\to& {v_a}' = (1-\f{p}{2}) v_a, \label{vwprime} \\
    w_{ab} &\to& {w_{ab}}' = (1-\f{3p}{2}) w_{ab} + \f{p}{4} \de_{ab} \, .
    \nn
\end{eqnarray}
\end{theorem}

\begin{proof}
The asymmetric quadruple product identity
\bel{quad}
    \sum_i J_i J_j J_k J_i = \delta_{jk} {\vec J}^{\, 2} - J_k J_j
\ee
implies the symmetrized identity
\bel{quad2}
    \sum_i J_i J_{(a} J_{b)} J_i
    =
    \delta_{ab} \vec J^2 - J_{(a} J_{b)} \ .
\ee
Using the latter and \er{tripleproduct}, a straightforward
calculation shows that $\Superop(\brho_{v,w})$ is equal to
\[
    \lrpBig{ 1 - \f{p}{2}}  v \cdot J
    +
    \lrpBig{ 1 - \f{3p}{2}} \sum_{a,b} w_{ab} J_{(a} J_{b)}
    + p \tr (w) \I \, ,
\]
which implies the stated result, since for spin-1, we have $\tr(w)
= 1/2$ and $\I = \sum_{a,b} \f12 \de_{ab} J_{(a} J_{b)}$.
\end{proof}

We refer to identities of the form \er{quad} as ``$4 \to 2$
identities,'' because they relate degree 4 polynomials in the
generators to degree 2 polynomials. We have also seen one ``$3 \to
1$ identity'' in equation \er{tripleproduct}.

It is possible to iterate formula \er{vwprime}, with interesting
results. Clearly, after $n$ applications of the channel, $v \to
(1-\f{p}{2})^n v$. Consider a W-state, i.e. a state of the form
\[
        \brho_{w} = \sum_{a,b} w_{ab} J_{(a} J_{b)} \, ,
\]
and let $\Superop^n$ denote $n$ applications of the spin-1
channel.

\begin{theorem}[Iteration Formula]
The action of $\Superop^n$ on $w$ is the following:
\[
    \xymatrix{ w \ar[r]_{\Superop^n} & \ }
    F^{(n)}(p) (\I - 6w) + w
\]
where $F^{(n)}(p)$ is a degree $n$ polynomial in $p$, determined
as follows. $F^{(1)}(p) = 1- 3p/2$, and the $F^{(n)}$ for $n > 1$
are determined by the recursion relation
\[
    F^{(n+1)}(p) = \lrpBig{1-\f{3p}{2}} F^{(n)}(p) + \f{p}{4} \, .
\]
\end{theorem}

Interestingly, this recursion relation has the same coefficients
as the transformation \er{vwprime} of $w$ itself.

\subsection{Pure States of the Spin-1 Channel} \label{sec:puresp1}

In any number of dimensions, one can find a class of pure states
in the following way. Let $\vec a \in \R^n$, and consider the
symmetric $n \times n$ matrix $P_{ij} = a_i a_j$. Then evidently,
\[
    P^2 = a^2 P \, ,
    \ \
    \te{ and }
    \ \
    \Tr(P) = a^2 \, .
\]
It follows that
\[
    P^2 = P \ \Leftrightarrow\  \Tr(P) = 1
    \ \Leftrightarrow\  \vec a \in S^n ,
\]
where $S^n$ denotes the $n$-dimensional sphere.

If $P$ is a density matrix, then it is a pure state. With $n = 3$,
these pure states are precisely the pure states that arise from
the symmetric term in \er{rhovw}, assuming we take the most
convenient choice of basis; i.e. the one in which the generators
for the spin-1 representation are
\[
  S_1 = \threemat{ 0 & 0 & 0 \\ 0 & 0 & -i  \\ 0 & i  & 0  }   ,\
  S_2 = \threemat{ 0 & 0 & i  \\ 0 & 0 & 0 \\ -i  & 0 & 0  }   ,\
  S_3 = \threemat{ 0 & -i  & 0 \\ i  & 0 & 0 \\ 0 & 0 & 0  } \, .
\]

Since density matrices of the form
\[
    \rho_w = \sum_{a,b} w_{ab} S_{(a} S_{b)} ,
\]
arise in the Bloch-type parameterization for spin-1, it is natural
to ask when this type of density matrix is pure. Solving the
equation
\bel{pure-eqn}
    (\rho_w)^2 = \rho_w
\ee
for the components of $w$, we find several two-parameter families
of solutions, and a one-parameter family of solutions. For the
two-parameter families, up to signs we have $\rho_w = P$, with the
$a$-vector given by
\[
    a_1 = \sqrt{w_{22} + w_{33}}, \ \
    a_2 = \sqrt{\textstyle \f12 - w_{22}}, \ \
    a_3 = \sqrt{\textstyle \f12 - w_{33}} \, .
\]
The off-diagonal coefficients of $w$ are also determined in terms
of $w_{22}$ and $w_{33}$, so there are indeed only two free
parameters.

More precisely, $\rho_w = P$ is one family of solutions; the
others are obtained by changing the signs of any two of the
off-diagonal components of $P$ above the diagonal, and changing
corresponding signs below the diagonal so that $P$ remains
symmetric. Hermiticity of $P$ requires that $\vec a$ must be a
real vector. This means that
\[
    w_{22} < \f12\,, \ \
    w_{33} < \f12\,, \ \
    w_{22} + w_{33} > 0\,,
\]
which constrains the point $(w_{22}, w_{33})$ to lie in the
interior of a certain triangle; see Figure 1.

\begin{center}
    \includegraphics[width=2in]{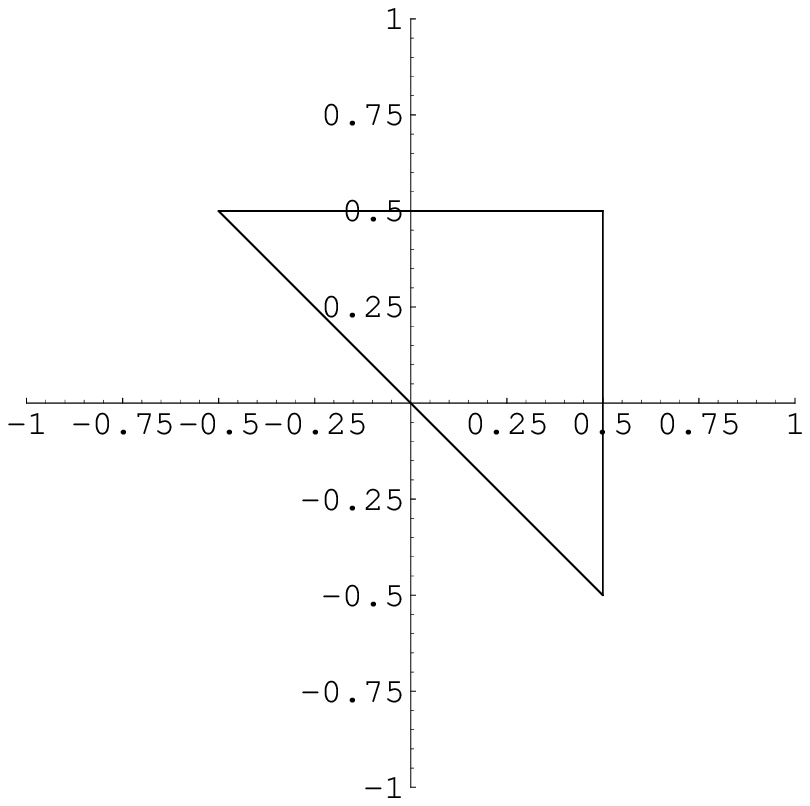}

    {\bf Figure 1.}
\end{center}

In addition to the two-parameter families discussed above, there
is a one-parameter family of solutions to \er{pure-eqn} taking the
form
\[
    P(\om) =
    \begin{pmatrix}
    \f12 + \om & 0 & \f12 \sqrt{1 - 4\om^2\phantom{{\big|}}}
    \\
    0 & 0 & 0
    \\
    \f12 \sqrt{1 - 4\om^2\phantom{{\big|}}} & 0 & \f12 - \om
    \end{pmatrix}
\]
where $\om = w_{33}$.

\subsection{Higher Spin Representations}

Note that the triple product \er{tripleproduct} and quadruple
product \er{quad} identities are simply certain elements of the
ideal $\mathscr{I} = \ker(\tilde{\phi})$, where $\tilde{\phi}$ is
the representation of the universal enveloping algebra, as in
\er{commdiag}. The larger this ideal, the more product identities
there will be in the representation of interest. For higher spin,
we have the following $3 \to 1$ identity in the spin $s$
representation of $\su_2$,
\bel{prod3_general_s}
    \sum_{i=1}^3 J_i J_a J_i = (\la -1) J_a \, ,
    \ \te{ where } \
    \la = s(s+1) \, .
\ee
There is also a generalization of the $4 \to 2$ identity valid for
general spin $s$,
\[
    \sum_{i=1}^3 J_i J_a J_b J_i = (\la - 2) J_a J_b
    +  \la \de_{ab} \I - J_b J_a \, .
\]
The latter has the more convenient symmetrized form:
\bel{prod4_general_s}
    \sum_{i=1}^3 J_i J_{(a} J_{b)} J_i = (\la - 3)
    J_{(a} J_{b)} + \la \de_{ab} \I \, .
\ee

\begin{theorem}[Higher Spin Channel] \label{thm:highspin}
Let $J_1, J_2, J_3$ be canonical generators for the spin-$s$
representation of $\su_2$ in dimension $d = 2s+1$, and let
$\Superop_s$ denote the spin-$s$ channel. Defining $\brho_{v,w} =
v\cdot J + \sum_{a,b} w_{ab} J_{(a} J_{b)}$, we have
\[
    \tr(\brho_{v,w}) = 1
    \ \Leftrightarrow \
    \tr(w) = \f{3}{d\la}
\]
where $\la = s(s+1)$. The action of the spin-$s$ channel is
\[
    \Superop_s(\rho_{v,w}) =
    \lrpBig{ 1 - \f{p}{\la} } v \cdot J
    +
    \lrpBig{ 1 - \f{3p}{\la} } w_{ab} J_{(a} J_{b)}
    + p \tr(w) \I \, .
\]
\end{theorem}

\begin{proof}
A straightforward application of \er{prod3_general_s} and
\er{prod4_general_s}.
\end{proof}

It is now clear that the action of the spin channel is more
complicated than the scaling of a single vector. It is the scaling
of a series of symmetric tensors, by different scale factors. This
shows that the spin-$s$ channels are never depolarizing channels.

At the critical value $p = \la / 3$, the channel maps an arbitrary
$\rho_{v,w}$ into a matrix with a Bloch representation:
\[
    \Superop_s(\brho_{v,w}) {\big|}_{p = \la/3}
    \ =\
    \f{1}{d} \I + \f{2}{3} v \cdot J \, .
\]
It follows that if $p = \la/3$, then the channel maps an initial
density matrix of the form \er{rhovw} with $v=0$ into pure noise.
We investigate critical values of $p$ more systematically in
Section \ref{sec:critical}.

For spin-1, an arbitrary density matrix may be represented as
\er{rhovw}, and for higher spin, these are a proper submanifold of
the convex cone of all density matrices. For spin 3/2, an arbitrary
density matrix may be written in the form
\[
    \sum_{a,b} w_{ab} J_{(a} J_{b)} +
    \sum_{a,b,c} u_{abc} J_{(a} J_b J_{c)}
\]
where $w$ and $u$ are completely symmetric tensors. The U-term is traceless,
and so we require the W-term to have trace one.
As discussed prior to Theorem \ref{thm:spin1}, this means that
$\tr(w) = 3/(d\la) = 1/5$.

\subsection{Finding $\bm{v}$ and $\bm{w}$ from $\bm{\rho}$
    in Higher Spin and Pure States for Spin One}

In this brief subsection we show how to invert the relation \er{rhovw}
for the density operator, and find the coefficient vector $v$ and symmetric
tensor $w$. We do the analysis at arbitrary spin, although for spin higher
than one, not all density matrices have the form \er{rhovw}. The methods
will generalize assuming the relevant trace identities can be found.

As in Theorem \ref{thm:highspin}, let $J_1, J_2, J_3$
be canonical generators for the spin-$s$ representation of
$\su_2$. Note that
\[
    \Tr( J_a J_b ) =
    \f13 d \la \, \de_{ab}
\]
where $\la = s(s+1)$ and $d = 2s+1$. Also,
\[
    \Tr(J_a J_b J_c) = i \, \f{d\la}{6} \, \eps_{abc}
\]
where $\eps_{abc}$ is the Levi-civita alternating symbol.
It follows immediately from \er{rhovw} that
\[
    v_a = \f{3}{d\la} \Tr(\rho J_a) \, .
\]
To find $w$, note the trace identity
\[
    \Tr( J_{(a} J_{b)} J_{(j} J_{k)} )
    =
    f_1(s) \f12 (\de_{ak} \de_{bj} + \de_{bk} \de_{aj})
    + f_2(s) \de_{ab} \de_{jk} \, ,
\]
where $f_i$ are functions of $s$, given by
\bea
    f_1 &=& \tr \lrp{ J_{(1} J_{2)} }^2
    = \f{\la d (d^2-4)}{30}  \\
    f_2 &=& \tr\lrp{ J_1^2 J_2^2 } = \f{\la d (1+2\la)}{30}\ .
\eea
By calculating
$\tr( \rho J_{(j} J_{k)} )$, we find
\bea
    w_{jk}
    &=&
    \f{1}{f_1} \lrpBig{ \tr\lrp{ \rho J_{(j} J_{k)} } -
        f_2\tr(w) \de_{jk} }  \\
    &=&
    \f{30}{\la d (d^2-4)} \tr\lrp{ \rho J_{(j} J_{k)} } -
        \f{2\la+1}{d^2-4} \tr(w) \de_{jk} \, .
\eea
For spin one, $f_1 = 1/2$, $f_2 = 1$, and $d\la = 6$ so
\[
    v_a = \f{1}{2} \tr(\rho J_a),
    \ \te{ and } \
    w_{jk} = \tr\lrp{ \rho J_{(j} J_{k)} } - \f12 \de_{jk} \, .
\]
This gives another way to find pure states: if $\rho = \ket{\psi} \bra{\psi}$ then
\bel{purestate-eqns}
    v_a = \f{1}{2} \<J_a\>_{\psi},
    \ \te{ and } \
    w_{jk} =
        \< J_{(j} J_{k)} \>_{\psi} - \f12 \de_{jk} \, .
\ee
The conclusion is that if the pure state $\rho = \ket{\psi} \bra{\psi}$
has a representation of the form \er{rhovw}, then we can find its
Bloch vector and $w$-matrix easily.

Using the same conventions as in Section \ref{sec:puresp1} for the
spin one operators, equation \er{purestate-eqns} implies that for
a pure state $\ket{\psi}$ with three complex components $\psi_a$,
we have
\[
    w_{ab} = \f12 \de_{ab} - \Re(\psi_a \bar{\psi_b}) \, .
\]
It also follows from \er{purestate-eqns} that
\bel{vecpsi}
    v = \vec\psi_R \times \vec\psi_I,
\ee
where $\vec\psi_R$ denotes the real vector with components
$\Re(\psi_a)$, and $\vec\psi_I$ has components $\Im(\psi_a)$. The
set of all $v$ satisfying \er{vecpsi} with $\< \psi \, | \, \psi\>
= 1$ is a ball of radius 1/2 in $\R^3$.

It seems that there are no pure states for spin 3/2 in the space
spanned by $J_a$ and $J_{(a} J_{b)}$. To find the pure states, it
is necessary to consider triples, i.e. $u$-states of the form
\[
    \sum_{abc} u_{abc} J_{(a} J_b J_{c)} \, .
\]

\subsection{Decomposition of the Space of Density Matrices into Convex Critical Regions}
\label{sec:critical}

In general, there is no value of $p \in [0,1]$ for which the
spin-1 channel maps all initial density matrices into pure noise,
so this channel is in some sense less decohering, and therefore
more desirable, than its spin-half counterpart. Rather, there are
two critical values, and two critical regions in the space $\cD$
of all density matrices.

Note that we may re-write \er{rhovw} in three dimensions as
\bel{new-rho_vw}
    \brho_{v,w} = \f13 \I + v \cdot J +
    \sum_{a,b} \lrpBig{w_{ab} - \f16 \de_{ab}} J_{(a} J_{b)} \, .
\ee
The term containing the symmetric tensor is now traceless, and
vanishes if and only if $w = \f16\, \I$. Define $R_0$ to be the
singleton set $R_0 = \{ \f13 \I \}$, and
\[
    \begin{matrix}
    R_1 = \{ v \cdot J : v \in \R^3 \}\qquad\qquad\qquad \\
    \ & \ \\
    R_2 = \Big\{ \sum_{a,b} \lrpBig{w_{ab} - \f16 \de_{ab}} J_{(a} J_{b)} \Big\}
    \end{matrix}
    \quad
    \begin{matrix}
        p_1 = 2 \\ \ \\ p_2 = \displaystyle \f{2}{3}
    \end{matrix}
\]
In other words, elements of $R_0, R_1, R_2$ are $3\times 3$
matrices that take the respective forms of the three terms in
\er{new-rho_vw}. Note that $R_1$ and $R_2$ are convex sets
containing only traceless matrices.

Moreover, any $3 \times 3$ density matrix can be written uniquely
as a sum of the form
\[
    r_0 + r_1 + r_2, \te{ where } r_i \in R_i .
\]
In compact notation, we have
\[
    \cD(\C^3) = R_0 + R_1 + R_2 \, , \quad
    R_i \cap R_j = \emptyset \ \te{ if } \ i \ne j,
\]
where $p_i$ is a \emph{critical value} for $R_i$, in the sense
that any $\rho \in R_0 + R_i$ is mapped to pure noise at $p =
p_i$. Density matrices not in $R_0 + R_i$ for some $i$ will not
have a critical value.

This kind of decomposition holds for some (but not all) Lie
algebras other than $\su_2$. The spin-1 example \er{new-rho_vw}
already well illustrates the fact that some of the critical values
of $p$ may lie outside the interval $[0,1]$ where the channel is
trace-preserving. In that case, only $p_2 = 2/3$ is a true
critical value, while $p_1 = 2$ does not describe a channel. In
our general discussion of this phenomenon, which culminates in
Theorem \ref{conj:critical-values}, we give an equivalent
condition for the critical values to lie in the allowed interval.

For any faithful, irreducible representation $\al$ of a semisimple
Lie algebra $\g$ on a Hilbert space $\cH$ with $d = \dim(\cH)$,
let $\Superop$ denote the associated channel \er{superop}. Let
$X_a$ denote generators of $\g$, and also their images under
$\al$. We use the term \emph{rank} to mean the degree of a
polynomial in $X_a$; for example $X_a X_b - \sum_i v_i X_a$ has
rank two. Rank is a well-defined function on the tensor algebra of
$\g$, but not on the universal enveloping algebra $\cU(\g)$, as
Lie algebra identities relate polynomials of differing ranks.
However, a given element of $\cU(\g)$ always has a (not
necessarily unique) representative of minimal degree in the
generators.

\begin{theorem} \label{conj:critical-values}
The space $\cD(\cH)$ of all density matrices on $\cH$ admits a
finite decomposition
\bea
    \cD(\cH) &=& R_0 + R_1 + \dots + R_N, \\
    && R_i \cap R_j = \emptyset \ \te{ if }\ i \ne j,
\eea
where $R_0 = \{ (1/d) \I \}$, each $R_r$ is a convex set
consisting of traceless degree $r$ combinations of the generators.
Further, $\exists\ p_r \in [0,1]$ such that
\[
    \Superop(\rho) = \f{1}{d} \I, \ \te{ at } \ p = p_r, \
    \te{ for all } \rho \in R_0 + R_r\,
\]
if and only if the generators in this representation satisfy
\emph{special $r \to r-2$ identities} with $g_r < 0$. (It is most
natural to define the term `special identity' and to define $g_r$
below, following eqn.~\er{specific-g}.)
\end{theorem}

\begin{proof}
Let $X_a$ denote the representation matrices in the representation
$\al$. By surjectivity of $\tilde{\al}$ in the commutative diagram
\er{commdiag}, we may write any density matrix $\rho$ as
\bel{rhodecomp}
    \rho = \sum_a v_a X_a + \sum_{a,b} w_{ab} X_{a} X_{b}
    + \sum_{a,b,c} u_{abc} X_{a} X_{b} X_{c} + \dots
\ee
Let $N$ be the smallest integer such that any $\rho \in \cD(\cH)$
can be written in the form \er{rhodecomp} with at most $N$ terms.

We may write any matrix $T$ as the sum of its trace part and its
trace-free part
\[
    T = T_{\tr} + T_0
\]
where
\[
    T_{\tr} = \f{1}{d}\, \Tr(T)\, \I,
    \ \te{ and }\
    T_0 = T - T_{\tr} \, .
\]
By assumption, \er{rhodecomp} has unit trace. It is then clear
that the sum of the trace part of each term must equal $(1/d) \I$.
We may therefore rewrite \er{rhodecomp} as
\[
    \rho = \f{1}{d} \I + \lrpBig{\sum_a v_a X_a}_0  +
    \lrpBig{\sum_{a,b} w_{ab} X_{a} X_{b}}_0 + \dots
\]
Define $R_0$ to be the singleton set $\{ \f1{d} \I\}$, and for $n
\geq 1$, define $R_n$ to be the set of all matrices of the form
\[
    \Big[
        \sum_{a_1, \ldots, a_n}
        w_{a_1, \ldots, a_n} X_{a_1} \dots X_{a_n}
    \Big]_0
\]
for all $w$ in the $n^{\te{th}}$ symmetric power of $\R^d$. It
follows that the space of all density matrices is decomposed as
\begin{eqnarray}
    \cD(\cH) &=& R_0 + R_1 + \dots + R_N, \label{DH-decomp} \\
    && R_i \cap R_j = \emptyset,\  i \ne j, \nonumber
\end{eqnarray}
Now let $\rho_r \in R_0 + R_r$, so that
\bel{rho_r}
    \rho_r = \f{1}{d} \I +
    \sum_{a_1, \ldots,a_r} \omega_{a_1 \dots a_r} X_{a_1} \ldots X_{a_r}
    - c\ \I
    \, ,
\ee
for some coefficient tensor $\omega$ and constant $c$ (equal to
the trace of the rank $r$ term).

To simplify notation, we describe the relevant procedure for a
rank 3 object $\rho_3 \in$ \mbox{$R_0 + R_3$}, with the
understanding that the generalization to arbitrary rank is
technically the same, but notationally worse. Writing
\[
    \rho_3 = \f{1}{d} \I + \lrpBig{\sum_{abc} \omega_{abc} X_a X_b X_c
    - c \I }
\]
where $c$ is a constant chosen to make the terms in parentheses
traceless, we then have
\bel{rho3ch}
    \Superop(\rho_3)
    =
    (1-p) \rho_3 + \f{p}{Z} \sum_{i=1}^k X_i \rho_3 X_i \, .
\ee
Recall that the representation-dependent constant $Z$ is defined
by the relation
\[
    \sum_i X_i^2 = Z \cdot \I
\]
and is related to the quadratic Casimir and the Killing-norm of
each of the generators.

Suppose that the representation being studied has a \mbox{$5 \to
3$} identity, so that
\bel{rtor-2}
    \sum_i X_i X_a X_b X_c X_i =
    f_{abc} \I + g_{abc} X_a X_b X_c
\ee
for some tensors $f,g$. There is \emph{no implied sum} on the rhs
of \er{rtor-2}. Consider using this to simplify \er{rho3ch},
keeping only the degree 3 terms in the generators. The result is
\[
    \sum_{abc} \omega_{abc} \lrpBig{
            1-p + \f{p}{Z} g_{abc}
        } X_a X_b X_c \, .
\]
To make this expression vanish, we would like to solve the
equation
\bel{specific-g}
    1-p + \f{p}{Z} g_{abc} = 0,
\ee
but that equation only yields a specific value for $p$ when
$g_{abc} = g_3 \in \C$ is a constant, i.e. takes the same
numerical value for any selection of the indices $a,b,c$.

This motivates the following definition of new terminology. In
general, when an $r \to r-2$ identity of the form \er{rtor-2}
holds with $g_{abc}$ equal to a \emph{scalar} $g_r \in \C$, let us
call it a \emph{special} $r \to r-2$ identity. We have proven
above that a special $4 \to 2$ identity (ie. $g_{abc}$ constant)
exists for any irreducible representation of $\su_2$, and found
the form of that identity. It is not hard to prove that for any
irrep of $\su_2$, special $n \to n-2$ identities exist for all
$n$. It is assumably an open question in representation theory
whether they exist for other representations; we hope that the
present work will motivate a further investigation of this
important question.

Assuming the special $5 \to 3$ identity in the example of
interest, we have
\begin{eqnarray}
    \Superop(\rho_3)
    &=&
    \lrp{\f{1}{d} + \f{p}{Z} \omega \cdot f - c } \I \label{pAwf} \\
    && +
    \lrb{ \lrp{\f{g_3}{Z} - 1} p + 1 } \sum_{abc} \omega_{abc} X_a X_b X_c
    \nonumber
\end{eqnarray}

Note that $\omega$ and $f$ are rank-3 tensors, and $\omega \cdot
f$ denotes the full contraction $\omega_{abc} f^{abc}$. The
coefficient of the rank 3 term vanishes at the value of $p$ which
sets the number in square brackets to zero. This critical value is
\[
    p_3 := \f{Z}{Z-g_3} \, .
\]
This value of $p$ is in the allowed range $[0,1]$ if and only if
$g_3 < 0$. The negativity of $g$ sometimes holds and sometimes
does not; for example, the 4$\to$ 2 identity \er{prod4_general_s}
would satisfy $g < 0$ for dimension less than 3. In any case, this
clarifies the point that a critical $p_r$ exists if and only if
there is a special $r \to r-2$ identity with $g_r < 0$.

Since at this value of $p$, the channel maps $\rho_3$ into $d^{-1}
\I$, we infer from \er{pAwf} that
\[
    \omega \cdot f = c (Z - g_3) \, .
\]
In the presence of a special $r \to r-2$ identity, the channel
maps $R_0 + R_r$ to itself, i.e.
\[
    \Superop(R_0 + R_r) \subset R_0 + R_r \, .
\]
This means that $\Superop$ effectively looks like a depolarizing
channel when restricted to $R_0 + R_r$. By choosing
\[
    p = p_r := \f{Z}{Z - g_r}
\]
only the term proportional to the identity survives. Since the
channel is trace-preserving, this term must be $d^{-1} \I$, and we
then have
\[
    \Superop(R_0 + R_r) \subset R_0 .
\]
\end{proof}

\subsection{Relation to the Werner-Holevo channel and a New Conjecture}

\citet{Datta04} has shown that the spin-1 channel at \mbox{$p =
1$} is equivalent to the Werner-Holevo channel
\bel{WHchannel}
    \Superop(\rho)
    =
    \f{1}{d-1} \lrp{ \tr(\rho) \I - \rho^T } \, .
\ee
Recall that in our notation, $M_0 = \displaystyle\sqrt{1-p}\ \I$,
so taking $p=1$ eliminates the identity from the set of Kraus
operators. For $p < 1$ and for the spin $s$ representation with $s
> 1$, we may view the spin channel as a generalization of the WH
channel.

The Werner-Holevo channel became famous as a counterexample to the
AHW conjecture \citep{AHW}. We infer by Datta's equivalence that
the spin-1 channel at $p=1$ gives precisely the same
counterexample to the AHW conjecture, stated below. Therefore,
multiplicativity does not hold generically in Lie algebra
channels. Once it was established that the AHW conjecture does not
hold for all  $q \geq 1$, it was natural to conjecture that it
holds for  $1 \leq q \leq 2$ \citep{KR}, and this was recently
proved for the WH channel by \citet{AF}. If this is true, one
would expect additional counterexamples with values of $q$
approaching $2$. However, none have yet been reported, except for
the WH channel which gives a sequence of counterexample with $q$
increasing from 4.79 as the dimension $d$ increase. M.~B.~Ruskai,
in a private communication to the author, suggested the
possibility that Lie algebra channels might provide additional
counterexamples with special properties:

\begin{conjecture}
Lie algebra channels generate counterexamples to the AHW
conjecture (stated below) for a sequence of values of $q$
approaching the boundary of the region in $q$-space where
multiplicativity begins to hold for all channels, assuming there
is such a region.
\end{conjecture}

For completeness, we now state the AHW conjecture, for which we
need a definition.

\begin{definition}
The \emph{maximal $\ell_q$-norm} of a channel $\Superop$ is
defined as
\[
    \nu_q(\Superop) = \sup_{\gamma  \in \cD(\cH)} \norm{\Superop(\gamma)}_q
    \quad
    (q \geq 1)
    \;,
\]
where $\norm{A}_q = \left( \Tr |A|^q \right)^{1/q}$, and
$\cD(\cH)$ denotes the space of density matrices on $\cH$.
\end{definition}

\citet{AHW} conjectured that $\nu_q(\Superop)$ is multiplicative
for tensor product channels:
\bel{mult}
    \nu_q(\Superop^{\otimes m})\equiv
    \sup_{\Gamma  \in \cD(\cH^{\otimes m})}
    \norm{\Superop^{\otimes m} (\Gamma)}_q =
    \nu_q(\Superop)^m \;.
\ee
Equation \er{mult} is often called the \emph{$\ell_q$
multiplicativity relation} or the \emph{AHW conjecture}.
\citet{GLR} have conjectured that \er{mult} holds for the
Werner-Holevo channel when $d \geq 2^{q-1}$.

\section{Channels Based on Exceptional Lie Algebras}
\label{sec:g2}

Let $e_j \ (j=0 \ldots 7)$ denote the standard basis for the
octonions $\bO$, where $e_0$ is the unit. Our notation is
compatible with that of \citet{Baez}, and the proofs of our
statements about the octonion algebra can be found there. The Lie
group $G_2$ is the automorphism group of $\bO$, so the Lie algebra
$\g_2$ is the derivation algebra of the octonions:
\[
    \g_2 = \der(\bO).
\]
Derivations act trivially on the identity, and the imaginary
octonions $\operatorname{Im}(\bO)$ form the fundamental
7-dimensional irreducible representation of $\g_2$.

It is known that if $\cA$ is an alternative, non-associative
algebra (such as the octonions), any pair of elements $x,y \in
\cA$ define a derivation $D(x,y) : \cA \to \cA$ by
\bel{D}
    D(x,y) a = [[x,y],a] - 3[x,y,a]
\ee
where $[a,b,x]$ denotes the associator $(ab)x - a(bx)$. When $\cA$
is a normed division algebra, every derivation is a linear
combination of derivations of this form. For the octonion algebra,
the elements
\[
    D(e_1,e_i), \ \
    D(e_2,e_j), \ \ \te{ and } \ \
    D(e_4,e_k)
\]
for all $i > 1, j > 2,$ and $k > 4$, are linearly independent and
there are 14 such elements, so they are a basis for $\g_2$. Define
the notation
\[
    d_{i,j} = \f12 D(e_i, e_j)
\]
This is one possible basis for the Lie algebra $\g_2$, but we will
use another more suited for our purposes. The fact
\citep{Macfarlane2002} that the six-dimensional sphere $S^6$ may
be viewed as a $G_2 / SU(3)$ coset space, implies a corresponding
decomposition of the algebra:
\[
    \g_2 = \frakm + \h, \quad \h \cong \su_3
\]
where $\frakm$ is a 6-dimensional subspace. We find a basis
adapted to this decomposition. The basis vectors for $\frakm$ are
simply expressed as $m_i = d_{1,i+1}$, while
\bea
  h_1 &=& d_{12} + 2 d_{47}, h_2 = d_{13} - 2 d_{46}, h_3 = d_{14} - 2 d_{27}, \\
  h_4 &=&
  d_{15} + 2 d_{26}, h_5 = d_{16} - 2 d_{25}, h_6 =
  d_{17} + 2 d_{24},\\
  h_7 &=&  \sqrt{3}\, d_{23}, h_8 = d_{23} + 2 d_{45}
\eea
are a basis for $\su_3$. Let
\[
    \beta = \f{i}{\sqrt{24}} \lrp{
        \{ m_1, \ldots, m_6\}
        \cup
        \f{1}{\sqrt{3}} \{ h_1, \ldots, h_8 \}
        }
\]
denote a corresponding basis for $\g_2$. Interestingly, $\beta$ is
an orthonormal basis of $\g_2$ with respect to the trace form on
the 7-dimensional representation space,
\[
    \Tr_{\operatorname{Im}(\bO)}(\beta_i \beta_j) = \f12 \delta_{ij}, \
    \te{ therefore } \
    \sum_{i=1}^{14} \beta_i^2 = I_7.
\]
The $\g_2$ channel acts as
\[
    \Superop(\brho) = (1-p) \brho + p
    \sum_{i=1}^{14}  \beta_i \,\brho\, \beta_i
    \, .
\]
Assume $\brho$ has a Bloch representation with $\vec v \in
\R^{14}$,
\bel{blochg2}
    \brho = \f17 \lrp{ I_7 + \vec v \cdot \vec \beta },
\ee
then as an intermediate step,
\[
    \Superop(\brho) = \f{1-p}{7} (I + \vec v \cdot \vec \beta)
     + \f{p}{7}
    \sum_{i=1}^{14}\lrp{ \beta_i^2 + v_a \beta_i \beta_a \beta_i }
    \, .
\]
The sum of $\beta_i^2$ gives the identity, with a factor of $p/7$
to cancel the $-p/7$, and (miraculously) the term which is cubic
in $\beta$'s vanishes identically! This is due to the following
remarkable $3 \to 0$ identity
\[
    \sum_i \beta_i \beta_a \beta_i = 0 \ \te{ for all } \ a,
\]
as may be checked explicitly. Therefore, the $\g_2$ channel
(restricted to its Bloch manifold) is the simplest of all. It is a
true depolarizing channel, shrinking its Bloch vector by a factor
of $1-p$,
\[
    \Superop(\brho) =
    \f{1}{7} \lrp{I + (1-p) \vec v \cdot \vec \beta} \, .
\]
We emphasize, however, that the $\g_2$ channel is almost certainly
\emph{not} a depolarizing channel outside the Bloch manifold,
though we have not proven this.

This does show that the critical value $p_1 = 1$, in the notation
of Theorem \ref{conj:critical-values}.

\section{Channels Based on the Clifford Algebra} \label{sec:cliff}

\newcommand\ga{\gamma}
\newcommand\n{\nu}
\def\Cl{\operatorname{C\ell}}

Let $\< \ , \ \>$ be a nondegenerate bilinear form on $V$, a
$d$-dimensional vector space. A representation of the Clifford
algebra associated to $(V, \< , \>)$ is a map $\ga : V \to \gl(V)$
satisfying
\[
    \{ \ga(x), \ga(y) \} = \< x,y \> \I \, .
\]
where the left side is an anticommutator. The representation is
\emph{Hermitian} if the image of $\ga$ is contained in $H(V)$, the
(Hilbert) space of Hermitian operators on $V$.

\begin{theorem}[Clifford Algebra Channel] \label{thm:cliff}
Given a Hermitian representation of the Clifford algebra, and a
finite collection of nonzero vectors $x_1, x_2, \ldots, x_n \in
\R^d$, then
\bel{cliff}
    \Superop_{\Cl}(\brho)
    \equiv
    \lrpBig{ \sum_{i=1}^n \< x_i, x_i \>}^{-1}
    \sum_{i=1}^n \ga(x_i)\, \brho\, \ga(x_i)
\ee
is a CPT map.
\end{theorem}

\begin{proof}
The operator is completely positive because it is already in the
form of an operator sum representation. We check that it is trace
preserving. By cyclicity of the trace,
\[
    \Tr( \Superop_{\Cl}(\brho) )
    =
    \lrpBig{ \sum_{i=1}^n \< x_i, x_i \>}^{-1}
    \sum_{i=1}^n \Tr( \brho\, \ga(x_i)^2 )
\]
However, $\ga(x_i)^2 = \f12 \{ \ga(x_i), \ga(x_i) \} = \< x_i, x_i
\> \I$ using the Clifford algebra. The sum of such terms decouples
from $\Tr(\brho)$ and exactly cancels the prefactor.
\end{proof}

We remark that, although the proof of Theorem \ref{thm:cliff} is
trivial, the result may not be easily obtained by inspecting any
of the standard matrix representations. Taking the Weyl
representation of the $\ga$ matrices in $d=4$,
one finds that writing out the CPT map
\[
    \ga(x) \brho \ga(x) + \ga(y) \brho \ga(y)
\]
for general $x,y,\brho$ as an explicit matrix takes a full page.

As we have seen in other examples, the computational methods used
in this paper are most effective when an arbitrary density matrix
can be written in terms of the generators of the symmetry algebra.
For the Weyl representation of the Clifford algebra, there is a
convenient basis consisting of antisymmetric combinations of $\ga$
matrices, which we summarize in the following table.
\begin{eqnarray*}
    \I & \te{ 1 of these} \\
    \ga^\m & \te{ 4 of these } \\
    \ga^{\m\n} = \f12 [\ga^\m, \ga^\n] = \ga^{[\m} \ga^{\n]}
    & \te{ 6 of these } \\
    \ga^{\m\n\rho} = \ga^{[\m} \ga^\n \ga^{\rho]}
    = i \epsilon^{\m\n\rho\sigma} \ga_\sigma \ga^5 &
    \te{ 4 of these } \\
    \ga^{\m\n\rho\sigma} = \ga^{[\m} \ga^\n \ga^{\rho} \ga^{\sigma]}
    = -i \epsilon^{\m\n\rho\sigma} \ga^5 &
    \te{ 1 of these }
\end{eqnarray*}
These 16 matrices form a basis for the space $\gl(\R^4)$. One can
therefore write any $4\times 4$ density matrix as a linear
combination of these matrices with coefficients that are tensors
of rank 4, and use $\ga$ matrix identities to calculate the action
of the CPT map \er{cliff}.

\section{The Bloch Manifold} \label{sec:blochmanifold}

\subsection{General Results}

The Bloch manifold is a geometrical space which is naturally
associated to a certain representation of a semisimple Lie algebra
$\g$, by asking the question: which linear combinations of the
generators of $\g$ in that representation can be valid density
matrices? For any preferred class of matrices (such as those with
nonnegative eigenvalues) one can define a manifold from a
representation in a similarly basis-dependent way, but for the
application to quantum physics, we restrict interest to density
matrices.

Why is this an important question? We have already shown that the
action of the $\su_n$ channel is most simply expressed as a
rescaling of the Bloch vector, and we know in that case that any
$n$-level density matrix admits a Bloch representation. All that
remains for a complete mathematical description of the $\su_n$
channel is to know the set of vectors $v \in \R^{n^2-1}$ on which
the transformation is being applied.

What about other representations? The simplest example of why this
is an important question for other representations is the formula
previously derived as \er{spin1b}, which gives the action of the
spin-1 channel on density matrices admitting a Bloch
representation ($\g = \su_2$) as:
\[
    \Superop(\brho_v) = \f13 \lrp{ \I + \big(1 - \f{p}2\big) v \cdot J}
    \, .
\]
Thus, the spin-1 channel also is a rescaling of the Bloch vector,
and so it is not only for the $\bm{n}$ representation of $\su_n$
that characterization of the Bloch manifold (the space of
admissible $v$) is important.

In any Lie algebra representation which has a ``triple product''
or $3 \to 1$ identity, i.e. an expression for $\sum_i \al(X_i)
\al(X_j) \al(X_i)$ in terms of the generators $\al(X_k)$, any
density matrix admitting a Bloch representation transforms very
simply under the action of the Lie algebra channel.

\begin{definition}
Choose a set of generators $\{ X_i \}$ of a semisimple Lie algebra
$\g$, and an irreducible representation $\al : \g \to \gl(\cH)$ on
a $d_\al$-dimensional Hilbert space $\cH$. The \emph{Bloch
manifold} $\cV$ (in the $X_i$ basis) is defined to be the set of
vectors $v \in \R^k \ (k = \dim \g)$, such that
\bel{bloch1}
    \bm{\rho}(v) = \f{1}{d_\al} \lrpBig{\I + \sum_i v_i \al(X_i)},
\ee
is a valid density matrix. A density matrix which can be written
in the form \er{bloch1} is said to possess a \emph{Bloch
representation}, and the corresponding vector $v$ is said to be a
\emph{valid Bloch vector}.
\end{definition}

In the notation of Theorem \ref{conj:critical-values}, the Bloch
manifold is precisely the space $R_0 + R_1$ which appears in the
natural decomposition of the space of all density matrices into
critical regions.

\begin{theorem}
The Bloch manifold is a closed set in $\R^k$.
\end{theorem}

\begin{proof}
The matrix $\bm{\rho}(v)$ is positive iff the lowest eigenvalue
$\la_{min}$  of $\I + \sum_i v_i \al(X_i)$ lies in the set $[0,
+\infty)$. The lowest eigenvalue of a matrix is a continuous
function of the matrix, so $\la_{min}$ is a continuous function of
$v$. The inverse image of the closed set $[0, +\infty)$ must be
closed.
\end{proof}

\begin{theorem} \label{thm:bloch_bound}
Let $\al$ be a $d$-dimensional representation of $\g$, let $k =
\dim(\g)$, and let $X_a$ be an orthogonal basis of $\g$ with
respect to the Killing form. Then
\[
    \Tr \lrp{ \al(X_a) \al(X_b) } = N d \, \de_{ab} \, .
\]
Moreover, if $v$ is a valid Bloch vector, then
\[
    v^2 \leq \f{d-1}{N} \ .
\]
\end{theorem}

We remark that in the notation of Section \ref{sec:basics},
\[
    N = Z / \dim(\g) \, .
\]

\begin{proof}
The density matrix $\brho = d^{-1} (\I + v_a \al(X_a))$ must
satisfy $\tr(\brho^2) \leq 1$. But
\bel{trace_rho_square}
    \tr(\brho^2) =
    d^{-1} (1 + N v^2)
\ee
which implies the desired result.
\end{proof}

If $\{X_j'\}$ is a second basis of $\g$, related to the original
basis by a matrix $A$, then the Bloch manifold in the $X'$ basis
consists of $A^T$ applied to the Bloch manifold in the $X$ basis.
If $\det(A) = 1$, this yields an isometric copy of the original
manifold, but otherwise the manifold has been stretched in some
way. We will see examples of Lie algebra representations which are
analogous to the qubit representation, in the sense that the Bloch
manifold is a closed ball in some preferred basis.

\subsection{The Bloch Manifold for All SU$\bm{{}_2}$ Representations.}

As an example, we give the Bloch manifold relevant to the
\mbox{spin-$j$} representation of $\su_2$. Let $I_{2j+1}$ be the
$(2j+1)$-dimensional identity matrix, and the $J_i$ are the
standard generators in the spin $j$ representation. The lowest
eigenvalue of
\[
    I_{2j+1} + \sum_{i=1}^3 v_i J_i
\]
is given by $1 - j\norm{v}$. We have proven:

\begin{theorem}
The valid Bloch vectors for the spin-$j$
representation of $\su_2$ (with the standard basis) are elements
of a closed ball in $\R^3$ with radius $1/j$.
\end{theorem}

Thus, the picture of the Bloch manifold as a closed ball is not
necessarily particular to the qubit system, however, it is
certainly not \emph{always} a closed ball. As we shall see below,
the Bloch manifold for the defining representation of $\su_n$ with
$n > 2$ is a proper subset of the analogous closed ball. To
complete the $\su_2$ case, we remark that the radius receives a
multiplicative constant if we rescale the generators; however, the
radius always scales as one inverse power of the dimension of the
representation.

\subsection{A Bloch Submanifold from the Cartan Subalgebra}

The $\al(X_i)$ are Hermitian operators on $\cH$, which cannot in
general be simultaneously diagonalized (if they can be, then
either $\g$ is abelian or the representation is trivial).
Therefore, solving the positivity condition in more sophisticated
examples is not straightforward. We discuss one method which works
for any Lie algebra representation and which always gives a
nonempty subset of the Bloch manifold.

Let $H_1, \ldots, H_r$ denote a basis for the Cartan subalgebra of
$\g$, with $r = \rank(\g)$. Simultaneously diagonalize all
$\al(H_i)$, and let ${h_i}^j$ denote the $j^{\text{th}}$ diagonal
element of $\al(H_i)$. The ${h_i}^j$ are, of course, weight
vectors for the given representation.

Assume that the basis $\{X_i\}$ has the Cartan generators $H_1,
\ldots, H_r$ as its first $r$ elements. We consider $v \in \R^k$
which are zero except for the first $r$ components, which
correspond to the Cartan generators, and ask when such a $v$ gives
rise to a density matrix. In this way we obtain a subset of the
Bloch manifold.

The positivity condition is
\[
    1 + \sum_{i=1}^{r} v_i {h_i}^j \geq 0
    \quad
    (\forall \ j = 1 \ldots d_\alpha) \, .
\]
Each linear equation $v \cdot h^j \geq -1$ defines a half-space
$\bH(j)$, and the restricted Bloch manifold
\[
    \cV_{\te{res}} =
    \{ v \mid v \cdot h^j \geq -1   \quad
    \forall \ j = 1, \ldots, d_\al \}
    = \bigcap_{j=1}^{d_\al} \bH(j)
\]
is their intersection, clearly nonempty. For representations of
nonabelian Lie algebras,
\[
    \cV_{\te{res}} \subsetneq \cV .
\]
Depending on the rank of $\g$, and on the spatial orientations of
the weight vectors, the space $\cV_{\te{res}}$ is either a finite
or a semi-infinite polyhedron.

\subsection{The Bloch Manifold for the Standard Rep of SU$\bm{{}_n}$.}

We now discuss the structure of the Bloch manifold for the
defining representation of $\su_n$. First, we derive a simple
bound by applying Theorem \ref{thm:bloch_bound} with $N = 2/n$,
which yields
\bel{sizeofv}
    v^2 \leq \f{n(n-1)}{2} \, .
\ee

By Descartes' rule of signs, an algebraic equation of degree $n$
with real roots,
\[
    \sum_{j} (-1)^j a_j x^{n-j} = \prod_{i=1}^{n} (x-x_i) = 0,
    \quad x_i \in \R
\]
has all roots nonnegative if and only if $a_i \geq 0$ for all $i$.
It is then obvious that the Bloch manifold for the $n$-dimensional
irrep of $\su_n$ is given by the set of \mbox{$v \in \R^{n^2-1}$}
such that the characteristic polynomial $\ch_{\brho(v)}(x)$ has
only nonnegative coefficients. The coefficients $a_i$ can be
calculated for any specific example using $a_0 = 1$ and Newton's
formula,
\bel{newton}
    a_k = \f{1}{k} \sum_{q=1}^k (-1)^{q-1} c_q a_{k-q},
\ee
where $c_q = \sum_i x_i^q = \tr\big( \brho(v)^q \big)$. Naturally,
calculating $\tr \brho^q$ reduces to calculating the traces of
products of at most $q$ generators of $\su_n$. Since
\[
    a_1 = c_1 = 1, \ \te{ and } \
    a_2 = \f12(1-c_2),
\]
the condition $a_2 \geq 0$ is equivalent to $\tr \brho^2 \leq 1$,
which leads to \er{sizeofv}.

Using \er{newton}, the condition $a_3 \geq 0$ reduces to $c_3 \geq
\f12 (3c_2 - 1)$, but $c_2$ is given by \er{trace_rho_square}, and
a similar calculation shows that
\[
    c_3 \equiv \tr \brho^3 =
    \f{1}{n^3} \lrp{ n + 6 v^2 + 2 v_a v_b v_c d_{abc} } \, .
\]

The calculations up to this point have been valid for $\su_n$ for
all $n$. However, to completely solve the problem for $n > 3$, we
will need to know $c_4, c_5, \ldots$ Therefore, as a nontrivial
example, we completely calculate the Bloch manifold for the
$\su_3$ channel in closed form. For $n = 3$, we note that
\[
    \det(v \cdot \la) = \f23 d_{ijk} v_i v_j v_k
\]
so the condition $c_3 \geq \f12 (3c_2 - 1)$ (for $n=3$) can be
expressed as
\[
    \det(v\la) > -1 \ \te{ and } \
    v^2 \leq 1 + \det(v\la)
\]
Therefore, the Bloch manifold for the $\bm{3}$ of $\su_3$ admits
the following expression, beautiful in its simplicity:
\[
    \cV_{\su_3} =
    \big\{
    v \in \R^8 \ : \
    v^2 \leq \min\lrpbig{3, 1 + \det(v\la) }
    \big\} \, .
\]

\subsection{Bloch Manifold for the $\bm{7}$ of G$\bm{{}_2}$.}

In our calculation of the $\g_2$ channel, we explicitly
constructed a basis $\beta$ of $\g_2$ using its definition as
$\der(\bO)$. This basis was normalized so that
\[
    \sum_a \beta_a^2 = \I, \ \
    \tr(\beta_a \beta_b) = \f12 \delta_{ab} .
\]
Theorem \ref{thm:bloch_bound} gives
\[
    |v| \leq 2 \sqrt{21} \, .
\]
This proves that the $\g_2$ Bloch manifold is contained in a
closed ball of radius about $9.2$. However, the true radius is
much smaller, as we will now show. By $\g_2$ symmetry, the
$\beta$'s satisfy the identity
\[
    \tr\ (v \beta)^q  = 0, \
    (\forall \ v \in \R^{14}), \
    q \ \te{ odd}
\]
where $v \beta = \sum_{i} v_i \beta_i$. Further, for certain even
values of $q$, $\tr\ (v \beta)^q$ may have a simple expression.
For example,
\begin{eqnarray}
    \label{vbeta4}
      \tr\ (v \beta)^2 &=& \f{v^2}{2}, \ \te{ and } \
      \\
      \tr\ (v \beta)^4 &=&
      \lrp{\tr\ (v \beta)^2}^2 = \f{v^4}{4} \, .
      \nn
\end{eqnarray}
These $\g_2$ trace identity $\tr\ (v \beta)^4 = \lrp{\tr\ (v
\beta)^2}^2$ is not easy to prove. It is true because for $\g_2$
and some other algebras, every fourth-order Casimir invariant is
expressible in terms of the second-order invariant, as shown by
\citet{Okubo}. Recently a simpler proof, together with other
interesting trace identities, was given by \citet[see
(4.36)]{Macf00}.

Enforcing $c_3 \geq \f12 (3c_2-1)$ gives a refinement,
\[
    |v| \leq 2\sqrt{7} \approx 5.3 .
\]
Requiring $a_4 \geq 0$ gives $v^2 \leq 8(10 - \sqrt{65})$, so $|v|
\leq 3.93$. The coefficients are such that $a_5 \geq 0$ for all
$v$, and \mbox{$\tr\ (v \beta)^n$} for $n \geq 6$ has no simple
expression analogous to \er{vbeta4}, so we have taken the simple
analysis of the $\g_2$ Bloch manifold as far as it will go.

\subsection{Pure States in the Bloch Manifold}

Let the representation matrices be denoted by $X_a$, $a = 1,
\ldots, k$. If $v$ is in the Bloch manifold, so that
\[
    \brho_v = d^{-1} (\I + \sum_a v_a X_a)
\]
is a density matrix, it is particularly easy
to determine whether $\brho$ is pure. If the products $X_a X_b$
are linearly independent from $X_a$ (i.e. there is no $2 \to 1$
identity) then $\brho^2 \ne \brho$ and the state is never pure.

On the other hand, if the representation has
a $2 \to 1$ identity of the type
satisfied by the fundamental representation of $\su_n$,
\bel{2to1}
    X_a X_b = \beta \delta_{ab} \I + \sum_c Q_{abc} X_c \, ,
\ee
then there can be pure states, and we have a complete
characterization of them.

\begin{theorem}[Pure Bloch States] \label{thm:pure}
If the $2\to 1$ identity \er{2to1} holds, then a Bloch state
$\brho_v$ is pure if and only if
\[
    1 + \beta v^2 = d \ \ \te{ and } \ \
     \sum_{a,b} v_a v_b Q_{abc} = \lrpBig{ 1 - \f{2}{d}} v_c \, ,
\]
for all $c = 1, \ldots, k$.
\end{theorem}

\noindent {\it Proof.} $\ $ This follows from
\[
    {\brho_v}^2 = \f{1}{d^2} (1 + \beta v^2) \I
    + \f{1}{d^2}
    \sum_{abc} ( 2 v_c + v_a v_b Q_{abc}) X_c\, . \quad \Box
\]

It is interesting to see how Theorem \ref{thm:pure} specializes to
$d = 2$. In that case, $Q_{abc} = 0$ and also $1 - 2/d = 0$, so
the second equation is always satisfied. The first equation
amounts to $v^2 = 1/ \beta$, and $\beta = 1$, so this just says
that $v$ is on the boundary of the Bloch sphere, which is the
well-known characterization of pure states.

Unfortunately, $2 \to 1$ identities almost never hold, excepting
of course the fundamental representation of $\su_n$, because
products $X_a X_b$ tend to be linearly independent from the
representation matrices $X_a$ if the dimension of the vector space
is large enough to allow this.

\subsection{Summary of Bloch Manifold Technology}

The Bloch manifold in a certain basis is given by the solution of
a system of polynomial inequalities in the components of the Bloch
vector $v$. These inequalities come from enforcing positivity of
the density matrix, $\brho_v \geq 0$. In many cases, it is easy to
see that the Bloch manifold is bounded within a ball, by enforcing
the inequality $\tr(\brho^2) \leq 1$. The Bloch manifold for the
$\bm{3}$ of $\su_3$ can be calculated exactly, and also in
principle for $\g_2$. In the latter case, it is bounded within a
ball of radius $< 3.93$. In any representation of any Lie algebra,
if a $2 \to 1$ identity \er{2to1} holds, then pure states lie on
the surface of a sphere of squared radius $(d-1)/\beta$.

What we have defined and studied here should rightly be called the
\emph{linear Bloch manifold}, because already for the spin-1
channel, one needs to represent the density matrix as $v\cdot J +
\sum_{a,b} w_{ab} J_{(a} J_{b)}$. So the full geometry of the
space of $3\times 3$ density matrices is described by placing
non-trivial conditions on both $v$ and $w$, and similar remarks
apply in higher dimensions.

For the spin 3/2 channel, one describes the most general density
matrix in terms of
\[
    \sum_{a,b} w_{ab} J_{(a} J_{b)} +
    \sum_{a,b,c} u_{abc} J_{(a} J_b J_{c)}
\]
where $w$ and $u$ are completely symmetric tensors. Thus the full
space of density matrices, in this representation, becomes a
submanifold of the space of ordered pairs $(w, u) \in V^{\otimes_s
2} \oplus V^{ \otimes_s 3}$ where $V$ is the 4-dimensional
representation space, satisfying some additional conditions. For a
general representation $V$, to generate all density matrices one
would need to consider a finite direct sum of symmetric tensor
powers $V^{\otimes_s n}$ for various $n$. The answer becomes more
complicated in higher dimensions (as does representation theory
itself) but the method is completely general.

\section{Conclusions}

We hope that the reader will find it useful to have a compilation
of results and formulae from the paper which lends itself to easy
reference for future research.

\begin{enumerate}
\item Definition of the generalized Lie algebra channel:
\[
    \Superop(\brho) = (1-p) \brho + \f{p}{Z}
    \sum_{i=1}^k \al(X_i) \,\brho\, \al(X_i)
\]
where the basis is orthogonal, and $Z$ is defined by $\sum_i
\al(X_i)^2 = Z\cdot \I$. If $X_i$ is Killing-orthonormal, then $Z$
is the quadratic Casimir.

\item  Extension yields many other channels, including
a ``double $\g$-channel'' with Kraus operators
\[
    \Big\{ \f{1}{Z} X_i X_j \ :\  i,j = 1\ldots k \Big\} \, .
\]
These operators generate the image of $\g \otimes \g$ under the
universal homomorphism expressed in the commutative diagram
\er{commdiag}.

\item Action of the $\su_n$ channel on an arbitrary input density
matrix:
\[
    \Superop(\brho_v)
    =
    \f1{n} \lrp{\I + \f{(1-p) n^2-1}{n^2-1}\, v \cdot X} ,
\]

\item Minimal von Neumann output entropy $S_{min}$ of the $\su_n$
channel:
\[
    \f{-np}{1+n}\,\ln \lrpBig{\f{np}{n^2-1}}  -
   \lrpBig{1 - \f{n\,p}{1 + n}} \ln \lrpBig{1 - \f{n\,p}{1 + n}}
\]
with large $n$ behavior:
\[
    \lim_{n \to \infty} \f{S_{min}}{\ln (n)} = p \, .
\]
This result was not discussed previously, but it is an easy
calculation.

\item An arbitrary $3 \times 3$ density matrix may be represented in the
form
\bel{last-rhovw}
    \brho_{v,w} = v\cdot J + \sum_{a,b} w_{ab} J_{(a} J_{b)}
\ee
for some vector $v$ and symmetric tensor $w$, with $\tr(w) = 1/2$,
though not all objects of this form are density matrices. The
action of the spin-1 channel on this density matrix is given by
scaling the vector and tensor according to
\bea
    v_a &\to& {v_a}' = (1-\f{p}{2}) v_a, \\
    w_{ab} &\to& {w_{ab}}' = (1-\f{3p}{2}) w_{ab} + \f{p}{4} \de_{ab} \, .
\eea

\item Iteration of the spin-1 channel $n$ times is the following
transformation on $w$:
\[
    \xymatrix{ w_{ab} \ar[r]_{\Superop^n} & \ }
    F^{(n)}(p) (\de_{ab} - 6w_{ab}) + w_{ab} \, ,
\]
where $F^{(n)}(p)$ is a degree $n$ polynomial in $p$. $F^{(1)}(p)
= 1- 3p/2$, and the $F^{(n)}$ for $n > 1$ are determined by the
recursion relation
\[
    F^{(n+1)}(p) = \lrpBig{1-\f{3p}{2}} F^{(n)}(p) + \f{p}{4} \, .
\]

\item Identities in the spin $s$ representation of $\su_2$:
\bea
    \sum_{i=1}^3 J_i J_a J_i &=& (\la -1) J_a \, ,
    \ \te{ where } \
    \la = s(s+1) \, \\
    \sum_{i=1}^3 J_i J_{(a} J_{b)} J_i &=& (\la - 3)
    J_{(a} J_{b)} + \la \de_{ab} \I
    \\
    \Tr( J_a J_b ) &=&
    \f13 d \la \, \de_{ab}
    \\
    \Tr(J_a J_b J_c) &=& i \, \f{d\la}{6} \, \eps_{abc}
    \\
    \Tr( J_{(a} J_{b)} J_{(j} J_{k)} )
    &=&
    f_1(s) \f12 (\de_{ak} \de_{bj} + \de_{bk} \de_{aj})
    + f_2(s) \de_{ab} \de_{jk} \, ,
\eea
where $f_i$ are functions of $s$, given by
\bea
    f_1 &=& \tr \lrp{ J_{(1} J_{2)} }^2
    = \f{\la d (d^2-4)}{30}  \\
    f_2 &=& \tr\lrp{ J_1^2 J_2^2 } = \f{\la d (1+2\la)}{30}\ .
\eea

\item
Let $J_1, J_2, J_3$ be canonical generators for the spin-$s$
representation of $\su_2$ in dimension $d = 2s+1$. For
$\brho_{v,w} = v\cdot J + \sum_{a,b} w_{ab} J_{(a} J_{b)}$ to be a
density matrix, we must have $\tr(w) = 3/(d\la)$ where $\la =
s(s+1)$. If $\Superop_s$ represents the spin-$s$ channel, then
\[
    \Superop_s(\rho_{v,w}) =
    \lrpBig{ 1 - \f{p}{\la} } v \cdot J
    +
    \lrpBig{ 1 - \f{3p}{\la} } w_{ab} J_{(a} J_{b)}
    + p \tr(w) \I \, .
\]

\item With $d$, $\la$, and $\rho_{v,w}$ as above, we have
\bea
    v_a &=& \f{3}{d\la} \tr\lrp{\rho_{v,w} J_a} \, ,
    \ \te{ and } \
    \\
    w_{jk}
    &=&
    \f{30}{\la d (d^2-4)} \tr\lrp{ \rho J_{(j} J_{k)} } -
        \f{2\la+1}{d^2-4} \tr(w) \de_{jk} \, .
\eea

\item If a density matrix $\rho_w$ is pure and takes the form
\[
    \rho_w = \sum_{a,b} w_{ab} S_{(a} S_{b)},
\]
(i.e. $v = 0$) then the matrix elements are $(\rho_w)_{ij} = \pm
a_i a_j$, where
\[
    a_1 = \sqrt{w_{22} + w_{33}}, \ \
    a_2 = \sqrt{\textstyle \f12 - w_{22}}, \ \
    a_3 = \sqrt{\textstyle \f12 - w_{33}} \, .
\]

\item
For the pure state $\psi$ with components $\psi_a$, we have
\[
    w_{ab} = \f12 \de_{ab} - \Re(\psi_a \bar{\psi_b}),
    \te{ and }
    v = \vec\psi_R \times \vec\psi_I,
\]
where $\vec\psi_R$ denotes the real vector with components
$\Re(\psi_a)$, and $\vec\psi_I$ for the imaginary part. The set of
all $v$ satisfying \er{vecpsi} with $\< \psi \, | \, \psi\> = 1$
is a ball of radius 1/2 in $\R^3$.

\item
The space $\cD(\cH)$ of all density matrices on $\cH$ admits a
finite decomposition
\bea
    \cD(\cH) &=& R_0 + R_1 + \dots + R_N, \\
    && R_i \cap R_j = \emptyset \ \te{ if }\ i \ne j,
\eea
where $R_0 = \{ (1/d) \I \}$, each $R_r$ is a convex set
consisting of traceless degree $r$ combinations of the generators.
Further, $\exists\ p_r \in [0,1]$ such that
\[
    \Superop(\rho) = \f{1}{d} \I, \ \te{ at } \ p = p_r, \
    \te{ for all } \rho \in R_0 + R_r\,
\]
if and only if the generators in this representation satisfy
\emph{special $r \to r-2$ identities} with $g_r < 0$.
\emph{Special identities} were defined following
eqn.~\er{specific-g}.

\item Action of the $\g_2$ channel on a Bloch-vector input:
\[
    \Superop_{\g_2}(\brho)
    =
    \f{1}{7} \lrp{ I + (1-p) \vec v \cdot \vec \beta} \, .
\]

\item When $\tilde{\phi}$, defined by the commutative diagram
\[
    \xymatrix{ \g \ar^i[r] \ar_{\phi}[dr] & \cU(\g) \ar^{\tilde{\phi}}[d] \\
    \ & \gl_d }
\]
is surjective, then the calculational technique used in this paper
will always work. This surjectivity holds under a very general set
of assumptions.

\item If $\ga$ denotes a representation of the Clifford algebra
associated to the bilinear form $\< \ , \ \>$ then the following
expression defines a channel:
\[
    \Superop_{\Cl}(\brho)
    \equiv
    \lrpBig{ \sum_{i=1}^n \< x_i, x_i \>}^{-1}
    \sum_{i=1}^n \ga(x_i)\, \brho\, \ga(x_i) \, .
\]

\item The \emph{Bloch manifold} is defined
to be the set of vectors $v \in \R^k$ such that
\[
    \bm{\rho}(v) = \f{1}{d_\al} \lrpBig{\I + \sum_i v_i \al(X_i)} \in
    \cD,
\]
It is always a nonempty closed set in $\R^k$, $k = \dim\g$.
Finding this manifold is important because it parameterizes the
space of density matrices for which the Lie algebra channel has a
simple formula.

\item In a $d$-dimensional representation of $\g$, normalized so
that $\Tr \lrp{ \al(X_a) \al(X_b) } = d N \de_{ab}$, for $v$ in
the Bloch manifold we have
\[
    v^2 \leq \f{d-1}{N}\ .
\]

\item In a certain basis, the Bloch manifold always
contains the intersection of the half-spaces $v \cdot h^j \geq
-1$, where $h^j$ are the weight vectors of the representation.

\item The valid Bloch vectors for the spin-$j$
representation of $\su_2$ (with the standard basis) are elements
of a closed ball in $\R^3$ with radius $1/j$.

\item The Bloch manifold for the $\bm{3}$ of $\su_3$ admits
the following beautiful expression:
\[
    \cV_{\su_3} =
    \big\{
    v \in \R^8 \ : \
    v^2 \leq \min\lrpbig{3, 1 + \det(v\la) }
    \big\} \, .
\]

\item Vectors $v$ in the Bloch manifold for the $\g_2$ channel,
with the chosen normalizations, satisfy
\[
    v^2 \leq 8(10 - \sqrt{65}), \ \te{ or } \ |v| \leq 3.93 \,.
\]

\item If the $2\to 1$ identity \er{2to1} holds, then $\brho$ is pure if
and only if
\[
    1 + \beta v^2 = d \ \ \te{ and } \ \
     \sum_{a,b} v_a v_b Q_{abc} = \lrpBig{ 1 - \f{2}{d}} v_c \, ,
\]
for all $c = 1, \ldots, k$.

\end{enumerate}

\section*{Acknowledgements}

I would like to thank Mary Beth Ruskai for helpful discussions,
John Preskill for his beautiful set of lecture notes on quantum
computation, and my Ph.D advisor Arthur Jaffe for support and
encouragement.

\bibliographystyle{plainnat}

\end{document}